\documentclass[reprint,floatfix,nofootinbib,superscriptaddress]{revtex4-2}

\usepackage[T1]{fontenc}
\usepackage{mathtools}
\usepackage{microtype}
\usepackage{color}
\usepackage[pass]{geometry}
\definecolor{bluesmoke}{rgb}{0.2,0.4,0.6}
\usepackage[
  allcolors=bluesmoke,
  colorlinks,
  pdftex,
  pdftitle={Thermal Fluctuations of Singular Bar-Joint Mechanisms},
  pdfauthor={Manu Mannattil, J. M. Schwarz, and Christian D. Santangelo},
  pdfkeywords={Classical statistical mechanics, Mechanical deformation, Potential energy surfaces, Origami \& Kirigami},
  pdfsubject={Statistical Physics, Polymers \& Soft Matter, Condensed Matter \& Materials Physics},
  hypertexnames=false
]{hyperref}
\usepackage{graphicx}
\graphicspath{{figures/}}

\usepackage[widespace]{fourier}
\DeclareMathAlphabet{\mathscr}{FMS}{futm}{m}{n}
\SetMathAlphabet{\mathscr}{bold}{FMS}{futm}{b}{n}
\DeclareMathAlphabet{\mathcal}{OMS}{lmsy}{m}{n}
\SetMathAlphabet{\mathcal}{bold}{OMS}{lmsy}{b}{n}
\usepackage[scale=0.9]{tgheros}
\DeclareMathAlphabet{\mathsf}{\encodingdefault}{\sfdefault}{\mddefault}{n}
\SetMathAlphabet{\mathsf}{bold}{\encodingdefault}{\sfdefault}{\bfdefault}{n}
\usepackage{bm}
\makeatletter
\def\big{\bBigg@{1}}
\def\Big{\bBigg@{1.5}}
\def\bigg{\bBigg@{2.4}}
\def\Bigg{\bBigg@{3.2}}
\makeatother

\makeatletter
\let\frontmatter@footnote@produce\frontmatter@footnote@produce@endnote
\def\bibsection{%
  \par
  \begingroup
  \baselineskip26\p@
  \bib@device{\hsize}{72\p@}%
  \endgroup
  \nobreak\@nobreaktrue
  \addvspace{19\p@}%
}%
\renewenvironment{acknowledgments}{}{}
\makeatother

\allowdisplaybreaks

\def\altsection#1{\paragraph*{#1.---\hskip-0.5em}}
\definecolor{hlcolor}{rgb}{0.68,0.1,0}

\DeclarePairedDelimiter{\abs}{\lvert}{\rvert}
\DeclarePairedDelimiter{\Abs}{\|}{\|}
\def\dd{\ensuremath\mathrm{d}}
\def\hess{\ensuremath\nabla\nabla}
\def\inmat#1{\ensuremath\begin{pmatrix}#1\end{pmatrix}}
\def\trans{\ensuremath{^\mathsf{T}}}

\begin{document}

\title{Thermal Fluctuations of Singular Bar-Joint Mechanisms}
\author{Manu Mannattil}
\email{mmannatt@syr.edu}
\affiliation{Department of Physics, Syracuse University, Syracuse, New York 13244, USA}
\author{J.~M.~Schwarz}
\email{jmschw02@syr.edu}
\affiliation{Department of Physics, Syracuse University, Syracuse, New York 13244, USA}
\affiliation{Indian Creek Farm, Ithaca, New York 14850, USA}
\author{Christian D.~Santangelo}
\email{cdsantan@syr.edu}
\affiliation{Department of Physics, Syracuse University, Syracuse, New York 13244, USA}

\begin{abstract}
  A bar-joint mechanism is a deformable assembly of freely-rotating joints connected by stiff bars.
  Here we develop a formalism to study the equilibration of common bar-joint mechanisms with a thermal bath.
  When the constraints in a mechanism cease to be linearly independent, singularities can appear in its shape space, which is the part of its configuration space after discarding rigid motions.
  We show that the free-energy landscape of a mechanism at low temperatures is dominated by the neighborhoods of points that correspond to these singularities.
  We consider two example mechanisms with shape-space singularities and find that they are more likely to be found in configurations near the singularities than others.
  These findings are expected to help improve the design of nanomechanisms for various applications.
\end{abstract}

\maketitle

\altsection{Introduction}

Bar-joint mechanisms constitute one of the simplest, widely-employed models to understand a variety of mechanical structures ranging from viruses~\cite{hespenheide2004},
colloidal clusters~\cite{holmes-cerfon2013,kallus2017,holmes-cerfon2017,robinson2019},
crystals~\cite{power2014} and minerals~\cite{kapko2011}, 
and robots and machines~\cite{farber2008,donelan2007}.
More recently, DNA origami has made the direct fabrication of miniaturized mechanisms possible at the nanoscale, where thermal fluctuations due to the surrounding medium cannot be neglected~\cite{marras2015,jung2020}.
As far as more generic descriptions of thermally-driven mechanisms are concerned, there has been long-standing interest in the effect of thermal fluctuations on the mechanical properties of ordered and disordered lattices~\cite{zhang2016,woodhouse2018,yan2018}, and the folding of polymerized membranes~\cite{di-francesco2000,nelson2004} and polyhedral nets~\cite{shenoy2012,dodd2018,melo2020}.
There is, therefore, an arising need to understand how thermal excitations affect the physical properties of these mechanisms, but only some attempts have been made so far~\cite{kallus2017,rocklin2018}.

\begin{figure}
  \begin{center}
    \includegraphics{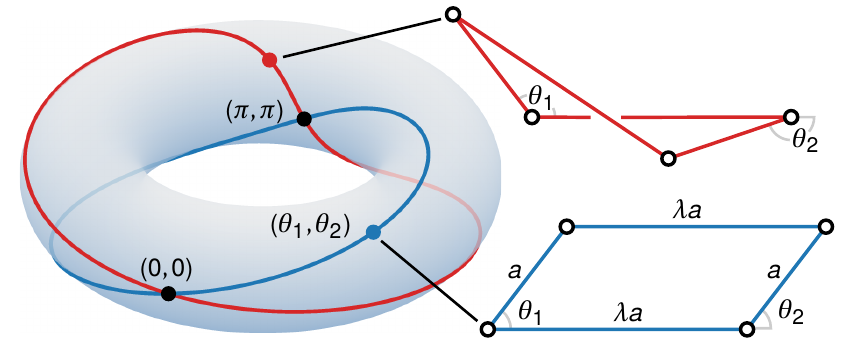}
  \end{center}
  \caption{Shape space of the planar four-bar linkage visualized as two intersecting curves on a torus, each curve representing a ``branch'' of the shape space.
    The poloidal and toroidal angles along the branches correspond to the angles $\theta_1$ and $\theta_2$ of the linkage, which has two modes of deformation with $\theta_{1} = \theta_{2}$ (blue curve) and $\theta_{1} \ne \theta_{2}$ (red curve).}
  \label{fig:4bar_cs}
\end{figure}

The effect of thermal fluctuations on a physical system is often represented by its free-energy landscape in terms of a set of collective variables that provide a coarse-grained description of its slowest dynamics.
In theory~\cite{go1976,echenique2011}, one can obtain the free energy of a mechanism by integrating out the fast modes that are transverse to its shape space, i.e., the subset of its configuration space once rigid-body motions are removed.
Doing this, however, becomes nontrivial when the mechanism has shape-space singularities~\cite{zlatanov2002,liu2003,donelan2007}.
For concreteness, consider the shape space of the planar four-bar linkage with freely rotating joints~\cite{grashof1883,hartenberg1964,shimamoto2005} (Fig.~\ref{fig:4bar_cs}).
Though this linkage has one degree of freedom up to Euclidean motions, it has two modes of deformation, one where the angle $\theta_1 = \theta_2$ and another where $\theta_1 \ne \theta_2$, meeting at two isolated singular points $(\theta_1,\theta_2) = (0,0)$ and $(\pi,\pi)$.
One generically expects the mechanism to be soft at these singularities, and indeed the free energy diverges in a harmonic approximation of the elastic energy~\cite{rocklin2018}.
These divergences must be cut off by higher-order nonlinear effects, yet how this happens and to what extent remains to be understood.

In this Letter, we develop a formalism to understand the thermal equilibration of common bar-joint mechanisms that have isolated shape-space singularities.
We show that the divergent contributions to the free energy arising in the harmonic approximation to the energy are suppressed by anharmonic corrections.
These findings show the existence of energetic free-energy barriers between configurations near the singularities and configurations farther from the singularities.
Our results are consistent with a closely-related work~\cite{kallus2017,holmes-cerfon2017} on singular colloidal clusters, but allow for isolated singularities of the shape space.
We demonstrate our results using both the four-bar linkage as well as a flat, triangulated origami~\cite{chen2018}.
Our analysis has direct consequences in the design and employment of nanoscale mechanisms in applications ranging from self-assembly~\cite{liedl2010} to drug delivery~\cite{zhao2019}, where relative thermodynamic stability of different configurations is of paramount importance.

\altsection{Mechanisms and singularities}

We consider bar-joint mechanisms made of $N \geq 3$ point-like joints in $d$ dimensions connected by $m < Nd - \frac{1}{2}d(d+1)$ freely-rotating, massless bars.
If the joints have position vectors $\bm{r}_1, \bm{r}_2, \ldots, \bm{r}_{N} \in \mathbb{R}^d$ in the lab frame, the mechanism's configuration can be fully described at any given moment using a configuration vector $\bm{r}\in\mathbb{R}^{N d}$ defined by $\bm{r} = (\bm{r}_1, \bm{r}_2, \ldots, \bm{r}_{N})$.
We assume the bars in the mechanism to be stiff but compressible with an energy that depends on the bar lengths so that the total energy of the mechanism is $U(\bm{r}) = \sum_{i=1}^{m} \phi_{i}[\ell_{i}(\bm{r})]$.
Here $\ell_{i}(\bm{r})$ is the length of the $i$th bar with an energy $\phi_i(\ell_{i})$, which is assumed to have a minimum value of zero at $\ell_{i} = \bar{\ell_{i}}$, the natural length of the $i$th bar.

With the above form of the energy, all nontrivial ground states of a mechanism belong to its shape space $\Sigma$ ~\cite{kendall1989,mezey1993,kendall1999}, which is the set of all deformed configurations of the mechanism with the length of each bar equal to its natural length, once rotations and translations are removed.
To practically identify $\Sigma$, we first switch to a Cartesian body frame attached to the mechanism so that all $\frac{1}{2}d(d+1)$ rigid motions are eliminated~\cite{herschbach1959,echenique2011}.
We require $n = Nd - \frac{1}{2}d(d+1)$ coordinates to specify the state of the mechanism in the body frame and let $\bm{q} \in \mathbb{R}^{n}$ be its configuration vector in this frame.
Now consider $m$ holonomic constraint functions $f_i: \mathbb{R}^{n} \to \mathbb{R},\,i=1,2,\ldots,m$, each associated with a single bar, and defined by $f_i(\bm{q}) = [\ell_i^2(\bm{q}) - \bar{\ell}_i^2]/(2\bar{\ell}_i)$.
The $m$ scalar constraint functions can also be considered together as a single constraint map $f: \mathbb{R}^{n} \to \mathbb{R}^m$ defined by $f(\bm{q}) = [f_1(\bm{q}), f_2(\bm{q}), \ldots, f_m(\bm{q})]$.
Then, the shape space is the zero level set $\Sigma = \left\{\bm{q} \in \mathbb{R}^{n}: f(\bm{q}) = \bm{0} \right\}$.
In the absence of external forces, each point in $\Sigma$ is a ground-state configuration of the mechanism with a distinct shape.

The compatibility matrix $\mathsf{C}(\bm{q})$~\cite{pellegrino1986,lubensky2015} at a configuration $\bm{q} \in \Sigma$ is the $m \times n$ Jacobian matrix $\nabla f$ of the constraint map $f$.
If $\mathsf{C}$ has full rank for all points in $\Sigma$, then $\Sigma$ is an $(n - m)$-dimensional submanifold of $\mathbb{R}^{n}$~\cite{leimkuhler2005,lee2013}.
When $\Sigma$ has a ``branched'' structure, e.g., like in Fig.~\ref{fig:4bar_cs}, $\mathsf{C}(\bm{q})$ drops rank at the singularity where the branches meet~\cite{lopez-custodio2020,muller2019}, and the constraints cease to be linearly independent.
Such singularities are the most common singularities~\cite{lopez-custodio2019,lopez-custodio2020} found in a mechanism and here we consider the situation where they occur only at isolated points of $\Sigma$.\footnote{For other, less common singularities that can occur in a mechanism, see Refs.~\cite{lopez-custodio2020,muller2019,muller2017}, and references therein.}
The branches of $\Sigma$, being $(n-m)$-dimensional submanifolds of $\mathbb{R}^{n}$, can be individually parameterized using a set of coordinates $\xi \in \mathbb{R}^{n-m}$, called shape coordinates~\cite{littlejohn1995} as they capture the shape changes of the mechanism as it moves on $\Sigma$.
We also assume that $n$ is small enough that such parameterizations can be found without much difficulty and that
the branches are linearly independent at the singularity~\cite{lopez-custodio2020}.
Zero-energy shape changes constitute the slowest dynamics in a mechanism, so it follows that the shape coordinates $\xi$ are the most natural collective variables (CVs) for a low-dimensional description of a thermally excited mechanism.

\altsection{Thermal fluctuations}

Let us assume that the value of the chosen CV for any configuration $\bm{q} \in \mathbb{R}^{n}$ of the mechanism can be measured using the CV map $\hat{\xi}(\bm{q})$.
(In the case of the four-bar linkage, for example, if we choose $\theta_{1}$ as the CV, then $\hat{\xi}(\bm{q)}$ is the map that computes $\theta_{1}$ for any $\bm{q}$, whether or not it lies on the branches of the linkage's shape space.)
The free energy associated with the CV $\xi$ is~\cite{lelievre2010}
\begin{equation}
  \mathcal{A}_{\hat{\xi}}(\xi) = -\beta^{-1}\ln{\mathcal{P}_{\hat{\xi}}(\xi)},
  \label{eq:free_energy}
\end{equation}
where $\beta$ is the inverse temperature and $\mathcal{P}_{\hat{\xi}}(\xi)$ is the marginal probability density of the CV, which, aside from factors of normalization, is
\begin{equation}
  \mathcal{P}_{\hat{\xi}}(\xi) = \int_{\mathbb{R}^{n}} \dd\bm{q}\, I(\bm{q})\, \delta\left[\hat{\xi}(\bm{q}) - \xi\right] \exp\left[-\beta U(\bm{q})\right].
  \label{eq:mpd}
\end{equation}
Here $\delta[\cdot]$ is the $(n-m)$-dimensional Dirac delta function, which restricts the domain of integration to the $m$-dimensional CV level set $\hat{\xi}^{-1}(\xi) = \left\{\bm{q} \in \mathbb{R}^{n}: \hat{\xi}(\bm{q}) = \xi\right\}$~\cite{hartmann2011}, and $I(\bm{q})$ is a Jacobian factor introduced by the change of coordinates from the lab frame to the body frame.
When $\hat{\xi}$ has full rank in $\hat{\xi}^{-1}(\xi)$, the coarea formula~\cite{lelievre2010} lets us express $\mathcal{P}_{\hat{\xi}}(\xi)$ as an exact high-dimensional surface integral over $\hat{\xi}^{-1}(\xi)$, but evaluating it is a difficult task in practice.
Hence, we have to resort to asymptotic methods for its evaluation.

At low temperatures (i.e., large $\beta$) we can asymptotically evaluate the integral in Eq.~\eqref{eq:mpd} by expanding the energy $U(\bm{q})$ around the ground-state configurations in $\hat{\xi}^{-1}(\xi)$.
Since all ground states belong to the shape space $\Sigma$, they could be regular (i.e., nonsingular) points or singularities of $\Sigma$.
We call $\xi$ a regular value of the CV if $\hat{\xi}^{-1}(\xi)$ does not contain singularities of $\Sigma$ and vice-versa.
For now, let us assume that $\xi$ is a regular value of the CV and that $\hat{\xi}^{-1}(\xi)$ contains just one ground state $\bar{\bm{q}}$.
If $\bm{q}$ is a point near $\bar{\bm{q}}$, after setting $\bm{q} \to \bar{\bm{q}} + \bm{q}$, we expand the energy to the lowest order around $\bar{\bm{q}}$ and find the harmonic energy $U \approx \frac{1}{2}\bm{q}\trans\mathsf{C}\trans\mathsf{K}\mathsf{C}\bm{q} = \frac{1}{2}\bm{q}\trans\mathsf{D}\bm{q}$.
Here $\mathsf{D} = \mathsf{C}\trans\mathsf{K}\mathsf{C}$ is the dynamical matrix evaluated at $\bar{\bm{q}}$~\cite{lubensky2015} (assuming joints of unit mass) and $\mathsf{K}$ is the diagonal matrix of bar stiffnesses $\phi_{i}''(\bar{\ell}_{i})$, which we set equal to $\kappa$ for all bars for simplicity.
See the Supplemental Material (SM)~\cite{supplemental} for details.
Since $\bar{\bm{q}}$ is a regular point of $\Sigma$, $\mathsf{C}$ has full rank, and $\mathsf{D}$ has $n-m$ independent zero modes that belong to $\ker\mathsf{C} = \left\{\bm{u} \in \mathbb{R}^{n}: \mathsf{C}\bm{u} = \bm{0}\right\}$~\cite{lubensky2015}.
These zero modes are all tangent to $\Sigma$ and represent a degree of freedom~\cite{leimkuhler2005}.
Hence, to asymptotically evaluate Eq.~\eqref{eq:mpd} in the neighborhood of a regular point, we can safely use the harmonic approximation since any divergence~\cite{schwarz1979,ellis1981,rocklin2018} due to these zero modes is regularized by the delta function, which suppresses all contributions to the integral that are tangent to $\Sigma$~\cite{ramond1997}.
Then, the asymptotic marginal density for a regular value $\xi$ of the CV is (SM~\cite{supplemental})
\begin{equation}
  \mathcal{P}_{\hat{\xi}}(\xi) \sim I(\xi)\left(\frac{2\pi}{\beta}\right)^{m/2}
  \left|\frac{\det\,[\nabla\psi(\xi)]\trans\nabla\psi(\xi)}{\det\,\mathsf{D}^{\perp}(\xi)}\right|^{1/2}.
  \label{eq:mpd_regular}
\end{equation}
Here $\psi: \mathbb{R}^{n-m} \to \mathbb{R}^{n}$ is a parameterization of $\Sigma$ near $\bar{\bm{q}} \in \Sigma$ in terms of the CV $\xi$, and compatible with the CV map, such that $\bar{\bm{q}} = \psi(\xi)$ and $\hat{\xi}(\bar{\bm{q}}) = \xi$.
Also, $\det\,(\nabla\psi)\trans\nabla\psi$ is the determinant of the induced metric on $\Sigma$ and $\mathsf{D}^{\perp}$ is the diagonal matrix of the $m$ nonzero eigenvalues of $\mathsf{D}$ at $\bar{\bm{q}}$.

Now, consider the situation at a shape-space singularity, where $\mathsf{C}$ has rank deficiency.
At such a point, using the Maxwell--Calladine count~\cite{maxwell1864,calladine1978}, we find that the number of zero modes increases to $n - m + s$, where $s$ is the number of independent self stresses $\bm{\sigma} \in \ker\mathsf{C}\trans$---each self stress being a set of bar tensions that leave the mechanism in equilibrium~\cite{lubensky2015}.
The zero modes at a singularity are not all tangent to $\Sigma$, which means that the delta function in Eq.~\eqref{eq:mpd} fails to suppress the divergences due to these zero modes when the harmonic approximation is used.
Furthermore, as one approaches the singularity along $\Sigma$, the lowest $s$ nonzero eigenvalues of the dynamical matrix $\mathsf{D}$ become small leading to an effective softening of the mechanism.
This causes Eq.~\eqref{eq:mpd_regular} to break down even for regular ground states in the vicinity of the singularity.
For instance, for the four-bar linkage, using Eq.~\eqref{eq:mpd_regular} we find $\mathcal{P}_{\hat{\theta}_1}(\theta_1) \sim |\sin \theta_1|^{-1}$ (SM~\cite{supplemental}), which diverges as $\theta_{1} \to 0, \pm\pi$.

To resolve the problem, we need to consider higher-order contributions to the energy due to the excess zero modes at the singularity.
Consider a singularity $\bar{\bm{q}}^{*} \in \Sigma$, where the CV has the value $\xi^{*}$.
For now, let us also assume that the only ground state in the CV level set $\hat{\xi}^{-1}(\xi^{*})$ is $\bar{\bm{q}}^{*}$.
For a point $\bm{q}$ close to $\bar{\bm{q}}^{*} \in \Sigma$, we set $\bm{q} \to \bar{\bm{q}}^{*} + \bm{q}$ and  write $\bm{q} = \bm{u} + \bm{v}$.
Here $\bm{u} \in \ker\mathsf{C}$ is a zero mode, $\bm{v} \in (\ker\mathsf{C})^{\perp}$ is a fast vibrational mode of the system, and $(\ker\mathsf{C})^{\perp}$ is the orthogonal complement of $\ker\mathsf{C}$ in $\mathbb{R}^{n}$.
Systematically expanding the energy to the lowest order in $\bm{u}$ and $\bm{v}$ around $\bar{\bm{q}}^{*}$~\cite{zhang2016,kallus2017,woodhouse2018} we find (SM~\cite{supplemental})
\begin{equation}
  U \approx \frac{1}{2}[\mathsf{C}\bm{v} + \bm{w}(\bm{u})]\trans\mathsf{K}[\mathsf{C}\bm{v} + \bm{w}(\bm{u})].
  \label{eq:energy_singular}
\end{equation}
Here $\bm{w}(\bm{u}) \in \mathbb{R}^{m}$ is a vector such that its $i$th component is $\frac{1}{2}\bm{u}\trans\hess f_{i}\bm{u}$, with $\hess f_{i}$ being the Hessian matrix of the $i$th constraint function $f_{i}$, evaluated at $\bar{\bm{q}}^{*}$.
This makes the above energy expansion quartic in the zero modes $\bm{u}$.

Equation~\eqref{eq:energy_singular} is only valid when the expansion is around the singularity $\bar{\bm{q}}^{*}$, and a similar expansion does not exist for ground states in $\hat{\xi}^{-1}(\xi)$ for $\xi$ close to $\xi^{*}$, where the harmonic approximation is not applicable either.
Thus, for $\xi \to \xi^{*}$, we choose to find $\mathcal{P}_{\hat{\xi}}(\xi)$ by directly evaluating the integral over $\hat{\xi}^{-1}(\xi)$ using the coarea formula.
To simplify the evaluation, we make two assumptions: (i) for points close to $\bar{\bm{q}}^{*}$, the CV map $\hat{\xi}$ can be approximated by its Taylor expansion around $\bar{\bm{q}}^{*}$: $\hat{\xi} = \xi^{*} + (\nabla\hat{\xi})\bm{q} + \mathcal{O}(\Abs{\bm{q}}^{2})$, with $\nabla\hat{\xi}$ being the Jacobian matrix of $\hat{\xi}$ at $\bar{\bm{q}}^{*}$; (ii) fast modes that belong to $(\ker\mathsf{C})^{\perp}$ do not change the value of the CV to linear order at $\bar{\bm{q}}^{*}$, i.e., $(\nabla\hat{\xi})\bm{v} = \bm{0}$.
Assumption (i) linearizes the CV map and turns its level sets near the singularity into hyperplanes, simplifying the evaluation of Eq.~\eqref{eq:mpd}.
Although assumption (ii) is stringent on the shape coordinate we use as the CV, it is true for most reasonable choices and a good CV should mainly be sensitive to the slow modes~\cite{tiwary2016}.
This makes it possible to use the quartic energy expansion and integrate over the fast modes.
Note that in the above steps, we do not make use of any parameterization of $\Sigma$, unlike in Eq.~\eqref{eq:mpd_regular}.

Using the linearized CV map and the quartic expansion for the energy [Eq.~\eqref{eq:energy_singular}] in Eq.~\eqref{eq:mpd}, we integrate out the fast vibrational modes $\bm{v}$ to find (SM~\cite{supplemental})
\begin{equation}
  \hspace{-0.65em}
  \thinmuskip=1mu \medmuskip=2mu 
  \begin{aligned}
    \mathcal{P}_{\hat{\xi}}(\xi) &\sim \frac{I({\xi}^{*})}{\left|\det\,\mathsf{D}^\perp \det\,\nabla\hat{\xi}(\nabla\hat{\xi})\trans\right|^{1/2}}  \left(\frac{2\pi}{\beta}\right)^{(m-s)/2} \\
                                 &\qquad \times \int_{\Xi_{\xi}}\hspace{-0.25em}\dd\Omega(\bm{u})\, \exp\left\{-\frac{1}{2}\beta\kappa\sum_{\bm{\sigma}}[\bm{\sigma}\cdot\bm{w}(\bm{u})]^2\right\},\enspace \xi \to {\xi}^{*},
  \end{aligned}
  \label{eq:mpd_singular}
\end{equation}
where $\bm{\sigma} \in \ker\mathsf{C}\trans$ are the self stresses and $\mathsf{D}^{\perp}$ is the diagonal matrix of the $m-s$ nonzero eigenvalues of $\mathsf{D}$ at $\bar{\bm{q}}^{*}$.
Also, $\dd\Omega(\bm{u})$ is the area element on the integration domain $\Xi_{\xi}$, which is geometrically an $s$-dimensional hyperplane formed by the intersection of $\ker\mathsf{C}$ and the level set of the linearized CV map $(\nabla\hat{\xi})^{-1}(\xi - {\xi}^{*}) = \{\bm{u} \in \mathbb{R}^{n} : \xi^{*} + (\nabla\hat{\xi})\bm{u} = \xi\}$.
On choosing a basis for $\ker\mathsf{C}$, the term in the exponential of the above integral becomes a quartic polynomial, making further simplification difficult.
We discuss the convergence criteria for Eq.~\eqref{eq:mpd_singular} in the SM~\cite{supplemental}.

On the basis of how $\mathcal{P}_{\hat{\xi}}(\xi)$ in Eqs.~\eqref{eq:mpd_regular} and \eqref{eq:mpd_singular} scales with $\beta$, we can show that the free-energy barriers between regular and singular values of the CV have a temperature/stiffness dependence $\sim\ln\beta\kappa$, making the barriers energetic in nature.
This is not surprising considering the overall softening of the mechanism near the singularities.
Also, for both the quartic and harmonic approximations for $\mathcal{P}_{\hat{\xi}}(\xi)$, we expect the range of validity (in $\xi$) to increase with increasing $\beta$, along with an increase in the range where both approximations produce similar results.
\begin{figure}
  \begin{center}
    \includegraphics{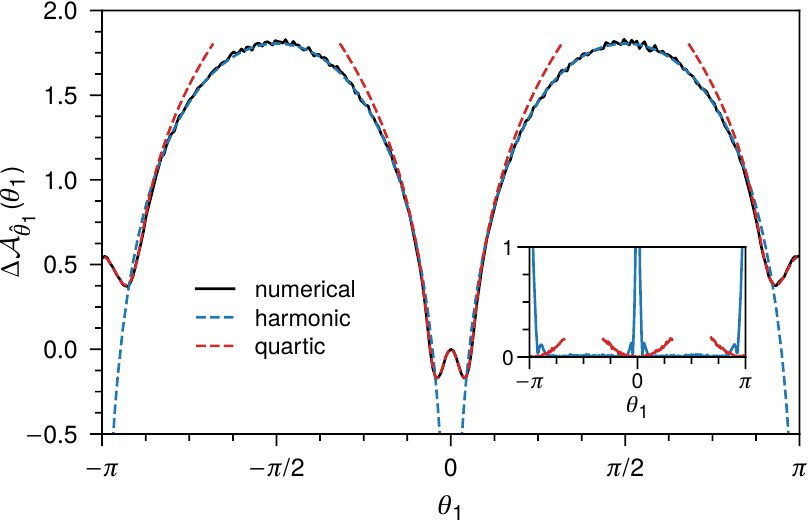}
  \end{center}
  \caption{Free-energy difference $\Delta\mathcal{A}_{\hat{\theta}_1}(\theta_1)$ of a four-bar linkage with parameters $a=1$ and $\lambda=2$ in units of $\beta^{-1}$ at $\beta = 10^{4}$.
  The inset shows the absolute errors between the numerical and asymptotic results using the harmonic approximation [Eq.~\eqref{eq:4bar_reg}, blue curve] and quartic approximation [Eq.~\eqref{eq:4bar_0}, red curves].}
  \label{fig:4bar_free}
\end{figure}

So far we have only considered cases where the CV level set $\hat{\xi}^{-1}(\xi)$ contains only one regular point or a singularity of $\Sigma$.
However, as $\Sigma$ has a branched structure, $\xi$ need not identify a configuration in $\Sigma$ uniquely.
Indeed, for the four-bar linkage, we see that there are as many as two configurations with a given value of $\theta_{1}$ (Fig.~\ref{fig:4bar_cs}).
Nonetheless, it is easy to find the asymptotic marginal density for more general cases by using combinations of Eqs.~\eqref{eq:mpd_regular} and \eqref{eq:mpd_singular} to add the contribution of each ground state in $\hat{\xi}^{-1}(\xi)$ individually, noting that Eq.~\eqref{eq:mpd_singular} gives the collective contribution from all the branches meeting at a singularity.
\begin{figure*}
  \begin{center}
    \includegraphics{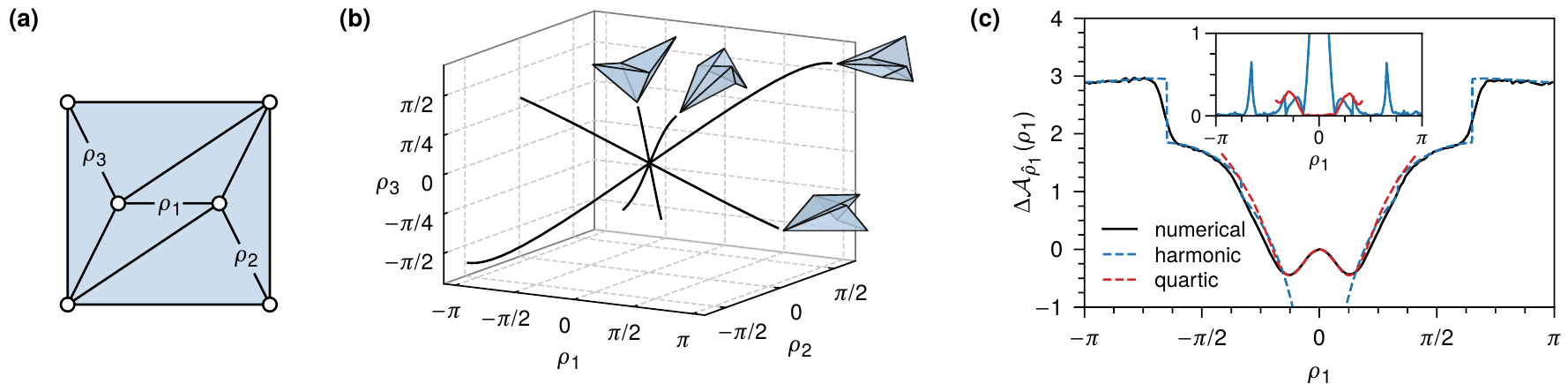}
  \end{center}
  \caption{(a) A triangulated origami modeled as a bar-joint mechanism and (b) its shape space visualized in the space of fold angles $\rho_{1}, \rho_{2}$, and $\rho_{3}$.
    (c) Free-energy difference $\Delta\mathcal{A}_{\hat{\rho}_1}(\rho_{1})$ in units of $\beta^{-1}$ at $\beta = 10^{4}$.
  The inset shows the absolute errors between the numerical and asymptotic results using the harmonic and quartic approximations (blue and red curves, respectively).}
  \label{fig:origami}
\end{figure*}

\altsection{Examples and discussion}

We now use our formalism to find the free-energy profiles of two example mechanisms with one-dimensional shape spaces with isolated singularities and compare them with results from Monte Carlo simulations.
(Also see the SM~\cite{supplemental} for an example mechanism with a two-dimensional shape space and a mechanism with a permanent state of self stress, which is unlike the case where it appears only at isolated singularities.)
Motivated by typical DNA origami structures that have lengths in the range of a few hundred nanometers with stiffness in the range 0.1--1~pN$/$nm~\cite{jung2020}, we choose a nondimensional inverse temperature of $\beta = 10^{4}$ and use a potential of the form $\phi_{i}(\ell_{i}) = (\ell_{i}^{2} - \bar{\ell}_{i}^{2})^{2}/(8\bar{\ell}_{i}^{2})$ so that $\phi_{i}''(\bar{\ell}_{i}) = \kappa = 1$.
Further details on the simulations are given in the SM~\cite{supplemental}.

The four-bar linkage we consider (Fig.~\ref{fig:4bar_cs}) is made out of two sets of bars of lengths $a$ and $\lambda a$, where $\lambda > 0$ is a dimensionless aspect ratio.
For $\lambda \ne 1$, the linkage has shape-space singularities at $\theta_{1} = 0$ and $\theta_{1} = \pm \pi$ where the bars become collinear and support a state of self stress.%
\footnote{For simplicity, we do not discuss the square four-bar linkage with $\lambda=1$ in this Letter as it has additional singularities at $(\theta_{1}, \theta_{2}) = (0, \pm\pi)$~\cite{yang1994}.}
The shape space can be fully parameterized using the angle $\theta_{1}$, which we use as our CV and choose $\theta_{1} = 0$ as the point of zero free energy.
For $\theta_{1}$ far from the singular values we use Eq.~\eqref{eq:mpd_regular} to find the free-energy difference $\Delta\mathcal{A}_{\hat{\theta}_1}(\theta_1) = \mathcal{A}_{\hat{\theta}_1}(\theta_1) - \mathcal{A}_{\hat{\theta}_1}(0)$ as~(SM~\cite{supplemental})
\begin{equation}
  \hspace{-0.5em}
  \Delta\mathcal{A}_{\hat{\theta}_1}(\theta_{1}) \sim \beta^{-1}\ln\left[X^{1/2}
    D_{-1/2}(0)|\sin{\theta_1}|\right],\enspace 0 \ll \abs{\theta_{1}} \ll \pi,
  \label{eq:4bar_reg}
\end{equation}
where $D_{-1/2}(\cdot)$ is the parabolic cylinder function~\cite{olver2010} and $X=\sqrt{\beta \kappa}\lambda a/(8|\lambda-1|)$ is a dimensionless term that is independent of $\theta_1$.
As expected, Eq.~\eqref{eq:4bar_reg} diverges when $\theta_{1}$ is close to the singular values $\theta_{1} = 0$ or $\theta_{1} = \pm\pi$.
For $\theta_{1} \to 0$, using Eq.~\eqref{eq:mpd_singular}, the free-energy difference takes the form~(SM~\cite{supplemental})
\begin{equation}
  \hspace{-0.5em}
  \Delta\mathcal{A}_{\hat{\theta}_1}(\theta_1) \sim \beta^{-1}\left\{X^2\theta_1^4 - \ln\left[\dfrac{D_{-1/2}(-2X\theta_1^2)}{D_{-1/2}(0)}\right]\right\},\enspace \theta_{1}\to 0\,.
  \label{eq:4bar_0}
\end{equation}
A similar expression is derived in the SM~\cite{supplemental} for $\theta \to \pm \pi$.
A comparison between the numerical results and asymptotic expressions in Eqs.~\eqref{eq:4bar_reg} and \eqref{eq:4bar_0} (Fig.~\ref{fig:4bar_free}) shows excellent agreement for all values of $\theta_{1}$

For further testing our methods, we consider an origami made by triangulating a unit square~\cite{chen2018} and embedded in three dimensions [Fig.~\ref{fig:origami}(a)].
To make the origami more realistic, in simulations, we avoid all configurations that result in face intersections.
The one-dimensional shape space [Fig.~\ref{fig:origami}(b)] of this origami can be visualized as four intersecting branches in the space of the fold angles, i.e., the supplement of the dihedral angle at a fold.
The intersection point is the singular flat state of the origami, where all the fold angles are zero.
After numerically parameterizing the branches of the shape space in terms of the fold angle $\rho_{1}$, which we use as our CV, we utilize Eq.~\eqref{eq:mpd_regular} to find the free energy $\mathcal{A}_{\hat{\rho}_{1}}(\rho_{1})$ for $\abs{\rho_{1}} \gg 0$.
We next find $\mathcal{A}_{\hat{\rho}_{1}}(\rho_{1})$ as $\rho_{1} \to 0$ using Eq.~\eqref{eq:mpd_singular} and choose $\rho_{1}=0$ as the point of zero free energy.
A comparison between the numerical and the asymptotic results for the free-energy difference  $\Delta\mathcal{A}_{\hat{\rho}_{1}}(\rho_{1})$ shows good agreement in both regimes of $\rho_{1}$ [Fig.~\ref{fig:origami}(c)].
Self-avoidance of the faces forces us to consider only a part of each branch of the shape space for our analysis.
Since the extent of these parts (in $\rho_{1}$) vary for the four branches [Fig.~\ref{fig:origami}(b)], it results in discontinuous jumps in the free-energy curves.

The free-energy landscapes of the four-bar linkage and the triangulated origami [Figs.~\ref{fig:4bar_free} and \ref{fig:origami}(c)] demonstrate that the measured values of the CV tend to be closer to their values near the singularities.
Yet, as free-energy landscapes (and even their extrema) do not always have a CV-agnostic interpretation~\cite{e2004,hartmann2007,frenkel2013}, to draw conclusions we should also consider the physical meaning of the chosen CV.
The CVs we picked for both the example mechanisms were internal angles whose values dictate the overall shape of the mechanism.
Specifically, according to our results, we expect the bars of the four-bar linkage to tend to be collinear, as measured by the angle $\theta_1$ being close to $0$ or $\pi$.
Similarly, the origami will tend towards being flat, as measured by the fold angle $\rho_1$.
This tendency increases at lower temperatures as the free-energy barriers become larger.
Finally, we remark on the apparent double-well nature of the landscapes near singular values of the CV.
Because of the branched nature of the shape spaces, when $\theta_{1} \to 0,\,\pm\pi$ or when $\rho_{1} \to 0$, there are multiple ground states where the mechanism is also relatively soft.
This is illustrated by the widening of the sublevel sets of the energy as one moves away from the singularity (e.g., see Fig.~\ref{sm:fig:energy34} of the SM~\cite{supplemental}).
The net result is an increase in the number of thermodynamically favorable states with $\theta_{1}$ close to $0$ or $\pm\pi$ (and $\rho_{1}$ close to $0$), causing an apparent lowering of the free energy.

\altsection{Conclusion}

In this Letter we have described a formalism to find the free-energy landscapes of common bar-joint mechanisms with isolated singularities in their shape spaces.
Our results indicate that configurations in the neighborhood of the singularities have relatively lower free energy compared to configurations farther from the singularities.
This could help in programming the conformational dynamics of nanomechanisms~\cite{dunn2015}.
Our findings also highlight the interplay between the geometry of a mechanism's shape space and its thermodynamic properties.
Since changes in configuration space topology is known to drive equilibrium phase transitions in certain physical systems~\cite{kastner2008}, it would then be interesting to consider how shape-space singularities affect the physical properties of mechanisms in the thermodynamic limit.
Indeed, the affinity for the origami to be in nearly-flat configurations is reminiscent of the well-known flat phase of a polymerized membrane~\cite{abraham1990,bowick1996}, which is the natural thermodynamic limit of the triangulated origami.
Other open questions include the behavior of these mechanisms in the presence of active (nonthermal) noise~\cite{woodhouse2018}, which is known to preferentially actuate zero modes, and methods to bias their dynamics towards desired states~\cite{kang2019}, e.g., by introducing CV-dependent bias potentials~\cite{kastner2011}.

\begin{acknowledgments}
  We acknowledge useful conversations with Xiaoming Mao and Zeb Rocklin.
  We also thank the anonymous reviewers for a careful reading of our manuscript.
  This work has been supported by the NSF Grants No.~DMR-1822638 and No.~DMR-1832002.
\end{acknowledgments}

%


\clearpage
\pagebreak

\newgeometry{width=400pt,height=620pt}
\makeatletter\def\set@footnotewidth{\onecolumngrid}\makeatother
\def\footnoterule{\hrule width 75pt\relax}
\setlength{\skip\footins}{\baselineskip}
\widetext

\setcounter{equation}{0}
\setcounter{figure}{0}
\setcounter{footnote}{0}
\setcounter{mpfootnote}{0}
\setcounter{page}{1}
\setcounter{paragraph}{0}
\setcounter{part}{0}
\setcounter{section}{0}
\setcounter{subparagraph}{0}
\setcounter{subsection}{0}
\setcounter{subsubsection}{0}
\setcounter{table}{0}

\def\theequation{S\arabic{equation}}
\def\thefigure{S\arabic{figure}}
\def\thepage{S\arabic{page}}

\begin{center}
  \large \textbf{Supplemental Material:\\Thermal Fluctuations of Singular Bar-Joint Mechanisms}\\[1.25em]
  \normalsize Manu Mannattil, J.~M.~Schwarz, and Christian D.~Santangelo\\
  \small \emph{Department of Physics, Syracuse University, Syracuse, New York 13244, USA}
\end{center}
\vspace{1.5em}

The organization of this Supplemental Material is as follows.
In Section~\ref{sm:sec:energy} we derive the lowest-order approximations of the energy [Eq.~\eqref{eq:energy_singular}].
In Sections~\ref{sm:sec:regular} and \ref{sm:sec:singular} we derive the asymptotic expressions for the marginal probability density for regular and singular values of the CV [Eqs.~\eqref{eq:mpd_regular} and \eqref{eq:mpd_singular}].
The planar four-bar linkage is discussed in Section~\ref{sm:sec:4bar}, including explicit parameterizations of the branches of its shape space as well as a detailed calculation of its free energy.
In Section~\ref{sm:sec:origami}, we discuss the triangulated origami and in Section~\ref{sm:sec:5bar}, we present results for the planar five-bar linkage---a mechanism with a two-dimensional shape space.
In Section~\ref{sm:sec:permanent}, we discuss mechanisms that are permanently singular, i.e., those with a permanent state of self stress.
Finally, in Section~\ref{sm:sec:numerics} we discuss the numerical simulations used in this work.

\section{Lowest-order Approximation of the Energy}
\label{sm:sec:energy}

The $i$th bar in the mechanism has a potential energy $\phi_i[\ell_i(\bm{q})]$.
The constraint function associated with this bar is $f_i(\bm{q}) = [\ell_i^2(\bm{q}) - \bar{\ell}_i^2]/(2\bar{\ell}_i)$.
Consider a point $\bar{\bm{q}}$ on the shape space $\Sigma$ where the mechanism is in equilibrium.
Dropping the index $i$ for now and noting that $\ell(\bar{\bm{q}}) = \bar{\ell}$, we first express the gradient and the Hessian of the length $\ell(\bm{q})$ at $\bm{q} = \bar{\bm{q}}$ in terms of the gradient and the Hessian of $f(\bm{q})$ as
\begin{equation}
\begin{aligned}
  \partial_j \ell &= \partial_j f,\\
  \partial_j\partial_k \ell &= \partial_j\partial_k f - \bar{\ell}^{-2}\partial_j f\partial_k f,
  \label{sm:eq:lentomap}
\end{aligned}
\end{equation}
where all the partial derivatives are with respect to the components of $\bm{q}$ and evaluated at $\bm{q} = \bar{\bm{q}}$.

Since we have assumed the potential energy of all bars to have a minimum value (assumed to be zero) on the shape space, where $\ell = \bar{\ell}$, the derivative $\phi'(\bar{\ell})$ vanishes.
We wish to find the lowest-order expansion of the total potential energy $U(\bm{q})$ near a point $\bar{\bm{q}} \in \Sigma$.
First, let us consider the case where $\bar{\bm{q}}$ is not a point of self stress, in which case all zero modes are tangent to $\Sigma$ and can be extended to smooth deformations of the mechanism that do not cost energy to any order.
In such a situation, after setting $\bm{q} \to \bar{\bm{q}} + \bm{q}$, the potential energy of a single bar to $\mathcal{O}(\Abs{\bm{q}}^2)$ is the familiar Hookean potential
\begin{equation}
  \phi[\ell(\bar{\bm{q}} + \bm{q})] =  \frac{1}{2}\phi''(\bar{\ell})(\partial_j \ell\,q_j)^2 + \mathcal{O}(\Abs{\bm{q}}^3).
\end{equation}
Above, the partial derivatives of $\ell(\bm{q})$ and the derivatives of $\phi(\ell)$ have been evaluated at $\bar{\bm{q}} \in \Sigma$ and $\ell = \bar{\ell}$, respectively.
The repeated indices are summed over as usual.
Reintroducing the free index $i$, and noting that the stiffness $\kappa_{i} = \phi_{i}''(\ell_{i})$ and $\partial_{j}\ell_{i}\,q_{j} = \partial_{j}f_{i}\,q_{j} = \nabla f_{i}\cdot\bm{q}$, the total energy becomes
\begin{equation}
  U = \sum_{i=1}^m \phi_i[\ell_i(\bar{\bm{q}} + \bm{q})] \approx \frac{1}{2} \sum_{i=1}^m \kappa_i (\nabla f_i\cdot\bm{q})^2 = \frac{1}{2}\bm{q}\trans\mathsf{C}\trans\mathsf{K}\mathsf{C}\bm{q} = \frac{1}{2}\bm{q}\trans\mathsf{D}\bm{q}.
  \label{sm:eq:regenergy}
\end{equation}

Now let us analyze the case where there are additional zero modes due to the presence of a self stress.
Dropping the index $i$ again, we first expand the energy of a bar to $\mathcal{O}(\Abs{\bm{q}}^4)$ and find
\begin{equation}
  \begin{aligned}
    \phi[\ell(\bar{\bm{q}} + \bm{q})] &=
    \frac{1}{2}\phi''(\bar{\ell})(\partial_j \ell\,q_j)^2 +
    \frac{1}{6}\phi'''(\bar{\ell})(\partial_j \ell\,q_j)^3 +
    \frac{1}{2}\phi''(\bar{\ell})(\partial_j \ell\,q_j)\partial_k\partial_l \ell\, q_k q_l\\ &\qquad+
    \frac{1}{24}\phi''''(\bar{\ell})(\partial_j \ell\,q_j)^4 +
    \frac{1}{4}\phi'''(\bar{\ell})(\partial_j \ell\,q_j)^2 \partial_k \partial_l \ell\, q_k q_l\\ &\qquad+
    \frac{1}{8}\phi''(\bar{\ell})(\partial_j\partial_k \ell\, q_j q_k)^2 +
    \frac{1}{6}\phi''(\bar{\ell})(\partial_j \ell\, q_j)\partial_k\partial_l\partial_m \ell\, q_k q_l q_m + \mathcal{O}(\Abs{\bm{q}}^5).
  \end{aligned}
  \label{sm:eq:energyexp}
\end{equation}
As in the main text, the subspace of zero modes is $\ker\mathsf{C}$ and its orthogonal compliment in $\mathbb{R}^n$ is $(\ker\mathsf{C})^\perp$ and we write $\bm{q} = \bm{u} + \bm{v}$, with $\bm{u} \in \ker\mathsf{C}$ and $\bm{v} \in (\ker\mathsf{C})^{\perp}$.
First let us analyze the contribution of the zero mode $\bm{u}$ to the energy.
Since $\bm{u}$ is a zero mode, we have $\mathsf{C}\bm{u} = \bm{0}$.
Now, as $\partial_j f$ is a row of $\mathsf{C}$, this implies $\partial_j f u_j = 0$.
Using this in Eq.~\eqref{sm:eq:lentomap} we get
\begin{equation}
  \begin{aligned}
    \partial_j \ell\,u_j &= \partial_j f u_j=0,\\
    \partial_j \partial_k \ell\, u_j u_k
                        &= \partial_j\partial_k f u_j u_k - \bar{\ell}^{-2}(\partial_j f\, u_j) (\partial_k f\, u_k)\\
                        &= \partial_j\partial_k f\, u_j u_k.
  \end{aligned}
\end{equation}
Setting $\bm{v} = \bm{0}$ in the series expansion [Eq.~\eqref{sm:eq:energyexp}] and using the above simplifications, it is clear that the only nonvanishing contribution to the energy would come from the $\mathcal{O}(\Abs{\bm{u}}^4)$ term $\frac{1}{8}{\phi''(\bar{\ell})}(\partial_j \partial_k \ell\, u_j u_k)^2 = \frac{1}{8}{\phi''(\bar{\ell})}(\partial_j \partial_k f\, u_j u_k)^2$.

To understand how a fast mode $\bm{v}$ contributes to the energy, we similarly set $\bm{u} = \bm{0}$ in the energy expansion.
Since $\bm{v}$ belongs to the orthogonal compliment of the subspace of zero modes, by definition, $\partial_j \ell v_j = \partial_j f v_j \ne 0$.
Hence, we see that the first nonvanishing contribution comes from the $\mathcal{O}(\Abs{\bm{v}}^2)$ term $\frac{1}{2}\phi''(\bar{\ell})(\partial_j \ell\, v_j)^2 = \frac{1}{2}\phi''(\bar{\ell})(\partial_j f\, v_j)^2$.
This shows that the energy scales as $\mathcal{O}(\Abs{\bm{u}}^4) \sim \mathcal{O}(\Abs{\bm{v}}^2)$, i.e., zero and fast modes respectively make quartic- and harmonic-order contributions to the energy.
Finally, we set $\bm{q} = \bm{u} + \bm{v}$ in Eq.~\eqref{sm:eq:energyexp}, and find the energy of the $i$th bar to the lowest order in $\bm{u}$ and $\bm{v}$ as
\begin{equation}
  \begin{aligned}
    \phi_i &= \frac{1}{2}\phi_i''(\bar{\ell}_i)(\partial_j\, f_i\, v_j)^2 + \frac{1}{2}\phi''_i(\bar{\ell}_i)(\partial_j f_i\, v_j)(\partial_k \partial_l f_i\, u_k u_l) + \frac{1}{8}\phi''_i(\bar{\ell}_i)(\partial_k \partial_l f_i\, u_k u_l)^2\\
           &\qquad+ \mathcal{O}(\Abs{\bm{u}}^5) + \mathcal{O}(\Abs{\bm{u}}^{3}\Abs{\bm{v}}) +\mathcal{O}(\Abs{\bm{u}}\Abs{\bm{v}}^{2}),
  \end{aligned}
\end{equation}
where we have reintroduced the free index $i$.
In the above equation, $\phi_{i}''(\bar{\ell}_i) = \kappa_i$, the stiffness of the $i$th bar, $\partial_j f_i v_j = \nabla f_i\cdot\bm{v}$, and $\partial_k\partial_l f_i\, u_k u_l = \bm{u}\trans(\hess f_i)\bm{u}$, with $\hess f_i$ being the Hessian matrix of the $i$th constraint function $f_i$.
Then, the energy of the $i$th bar to the lowest order is
\begin{equation}
  \phi_i \approx \frac{1}{2}\kappa_i \left[\nabla f_i\cdot\bm{v} + \frac{1}{2}\bm{u}\trans(\hess f_i)\bm{u}\right]^2.
\end{equation}
Recognizing that $\frac{1}{2}\bm{u}\trans(\hess f_{i})\bm{u}$ is the $i$th component of the column vector $\bm{w}(\bm{u}) \in \mathbb{R}^{m}$ defined in the main text, the total energy to the lowest order in $\bm{u}$ and $\bm{v}$ takes the form
\begin{equation}
  U = \sum_{i=1}^m \phi_i \approx \frac{1}{2} [\mathsf{C}\bm{v} + \bm{w}(\bm{u})]\trans\,\mathsf{K}\,[\mathsf{C}\bm{v} + \bm{w}(\bm{u})].
\end{equation}
This equation is a generalization of the energy of harmonic spring networks~\cite{zhang2016,woodhouse2018} and colloidal clusters~\cite{kallus2017} that have zero modes, for an arbitrary number of zero modes and degrees of freedom, as well as a general interaction energy $\phi_{i}$ between the particles (i.e., the joints).

\section{Asymptotic Expressions for the Marginal Probability Density}

  As we remarked in the main text, the presence of the delta function restricts the domain of integration in the marginal density $\mathcal{P}_{\hat{\xi}}(\xi)$ [Eq.~\eqref{eq:mpd}] to the CV level set $\hat{\xi}^{-1}(\xi) = \left\{\bm{q} \in \mathbb{R}^{n}: \hat{\xi}(\bm{q}) = \xi\right\}$~\cite{hartmann2011}.
Now, since the CV map is $\hat{\xi}: \mathbb{R}^{n} \to \mathbb{R}^{n-m}$, the CV level set $\hat{\xi}^{-1}(\xi)$ will be an $m$-dimensional manifold if $\nabla\hat{\xi}$ has full rank in $\hat{\xi}^{-1}(\xi)$.
Then, the marginal density $\mathcal{P}_{\hat{\xi}}(\xi)$ can be written as an exact $m$-dimensional surface integral over $\hat{\xi}^{-1}(\xi)$ using the coarea formula~\cite{hartmann2007,lelievre2010,hartmann2011,diaconis2013},
\begin{equation}
  \mathcal{P}_{\hat{\xi}}(\xi) = \int_{\hat{\xi}^{-1}(\xi)} \frac{\dd\Omega(\bm{q})}{\abs{\det\,\nabla\hat{\xi}(\nabla\hat{\xi})\trans}^{1/2}} I(\bm{q}) \exp{\left[-\beta U(\bm{q})\right]}.
  \label{sm:eq:coarea}
\end{equation}
Here $\dd\Omega(\bm{q})$ is the surface measure on $\hat{\xi}^{-1}(\xi)$ and $\nabla\hat{\xi}$ is the $(n-m)\times n$ Jacobian matrix of $\hat{\xi}$ at $\bm{q}$.
Sometimes, Eq.~\eqref{sm:eq:coarea} is taken to be the definition of $\mathcal{P}_{\hat{\xi}}(\xi)$ instead of Eq.~\eqref{eq:mpd}.
This only makes sense when $\nabla\hat{\xi}$ has full rank in $\hat{\xi}^{-1}(\xi)$ so that the determinant $\det\,\nabla\hat{\xi}(\nabla\hat{\xi})\trans$ does not vanish~\cite{lelievre2010}.%
\footnote{Equation~\eqref{sm:eq:coarea} is the arbitrary-dimensional analogue of the RHS in $\int_{\mathbb{R}} \dd{x}\, f(x)\, \delta[g(x)] = \sum_{i} f(x_{i})/\abs{g'(x_{i})}$, where $f$ and $g$ are scalar functions in $\mathbb{R}$, and $x_{i}$ are the roots of $g(x) = 0$. This is assuming $\abs{g'(x_{i})} \neq 0$, which is equivalent to the assumption that $\nabla\hat{\xi}$ has full rank in $\hat{\xi}^{-1}(\xi)$.}
Furthermore, although we have in mind an $(n-m)$-dimensional CV $\xi$ that can be used to parameterize the branches of the shape space $\Sigma$, we do not require a parameterization at hand to use Eq.~\eqref{sm:eq:coarea}.
For instance, in an origami the CV could be one of its fold angles, e.g., $\rho_{1}$ in Fig.~\ref{fig:origami}, whose value can be directly computed from the coordinates of the origami.
(The map that turns the coordinates $\bm{q}$ into the fold angle is the CV map $\hat{\xi}(\bm{q})$ for the origami.)
Hence, in theory, using this equation only requires knowledge of the energy $U(\bm{q})$, the Jacobian factor $I(\bm{q})$, and the CV map $\hat{\xi}$.
Most importantly, Eq.~\eqref{sm:eq:coarea} makes no reference to the shape space $\Sigma$, or its branches, or whether or not it has singularities.
However, in general, the CV level set $\hat{\xi}^{-1}(\xi)$ is bound to be a curved high-dimensional manifold.
Hence, directly evaluating the integral in Eq.~\eqref{sm:eq:coarea} becomes cumbersome, and we have to resort to asymptotic methods to evaluate it.

It is clear from both Eqs.~\eqref{eq:mpd} and \eqref{sm:eq:coarea} that in the large-$\beta$ limit, contributions to the marginal density would mainly come from the neighborhoods of the ground states in the CV level set $\hat{\xi}^{-1}(\xi)$ since the energy $U(\bm{q})$ vanishes at those points (see Fig.~\ref{sm:fig:levelsets}).
Therefore, after asymptotically evaluating the integral in Eq.~\eqref{eq:mpd} in the neighborhood of each ground state (e.g., using Laplace's method~\cite{breitung1994}), we can then sum the results to find the asymptotic expression of the marginal density.
In the following, to simplify the presentation, we will only consider cases where the CV level set contains just one ground state, which is either a regular point or a singularity of $\Sigma$.
As we mentioned in the main text, more general cases can then be handled by using appropriate combinations of the results we derive.

\subsection{Regular values}
\label{sm:sec:regular}

We first consider the case where $\xi$ is a regular value of the CV, i.e., when the CV level set $\hat{\xi}^{-1}(\xi)$ contains only regular points of $\Sigma$.
As we remarked previously, using Eq.~\eqref{sm:eq:coarea} to find the marginal density is difficult.
Hence, we will use a Gaussian representation of the delta function to write the marginal density as~\cite{hartmann2007a,hartmann2011}
\begin{equation}
  \mathcal{P}_{\hat{\xi}}(\xi) = \lim_{\alpha\to\infty} \left(\frac{\alpha}{2\pi}\right)^{(n-m)/2} \int_{\mathbb{R}^n} \dd \bm{q}\, I(\bm{q}) \exp{\left[-\frac{1}{2}\alpha\Abs{\hat{\xi}(\bm{q}) - \xi}^2 - \beta U(\bm{q})\right]},
  \label{sm:eq:prob_delta}
\end{equation}
where $\Abs{\cdot}$ is the $(n-m)$-dimensional Euclidean norm.
For $\bm{q} \in \hat{\xi}^{-1}(\xi)$, the norm $\Abs{\hat{\xi}(\bm{q}) - \xi}$ vanishes.
Similarly, for $\bm{q} \in \Sigma$, the energy $U(\bm{q})$ vanishes.
This means that in the limit $\alpha, \beta \to \infty$, contributions to the above integral would mainly come from the neighborhood of the ground state $\bar{\bm{q}} = \Sigma \cap \hat{\xi}^{-1}(\xi)$.
Hence, we can evaluate the above integral using Laplace's method after expanding the two terms in the exponent of Eq.~\eqref{sm:eq:prob_delta} to the lowest order around $\bar{\bm{q}}$.
This gives us
\begin{equation}
  \mathcal{P}_{\hat{\xi}}(\xi) \sim \lim_{\alpha\to\infty} \frac{\alpha^{-m/2}}{(2\pi)^{(n-m)/2}}I(\bar{\bm{q}})\int_{\mathbb{R}^n} \dd{\bm{q}}\, \exp\left\{-\frac{1}{2}\bm{q}\trans\left[(\nabla\hat{\xi})\trans\nabla\hat{\xi} + \alpha^{-1}\beta \mathsf{D}\right]\bm{q}\right\},
  \label{sm:eq:prob_gaussian}
\end{equation}
where $\mathsf{D}$ is the dynamical matrix [see Eq.~\eqref{sm:eq:regenergy}] and $\nabla\hat{\xi}$ is the Jacobian matrix of $\hat{\xi}$, both evaluated at $\bar{\bm{q}}$.
We have also rescaled $\bm{q} \to \alpha^{-1/2}\bm{q}$ for convenience.
\begin{figure}
  \begin{center}
    \includegraphics{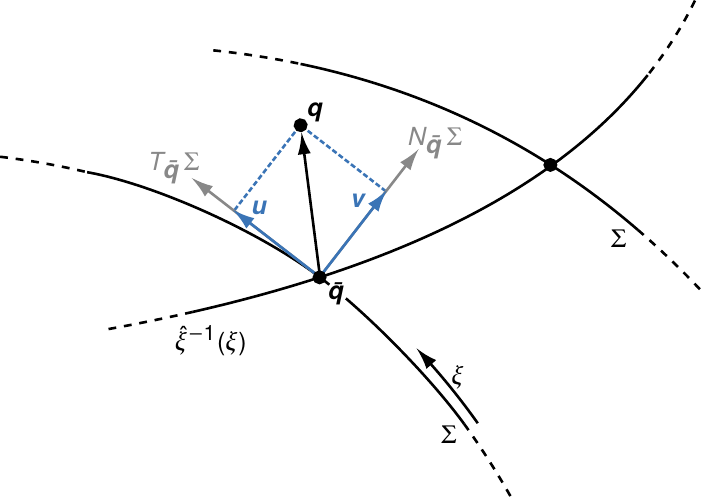}
  \end{center}
  \caption{Cartoon illustrating the shape space $\Sigma$ parameterized by the CV $\xi$ and the CV level set $\hat{\xi}^{-1}(\xi)$.
    In general, the sets $\Sigma$ and $\hat{\xi}^{-1}(\xi)$ could intersect at multiple points, each of which is an energy ground state.
    Tangent and normal spaces at such a point $\bar{\bm{q}} \in \Sigma \cap \hat{\xi}^{-1}(\xi)$ are $T_{\bar{\bm{q}}}\Sigma$ and $N_{\bar{\bm{q}}}\Sigma$.}
  \label{sm:fig:levelsets}
\end{figure}

Consider evaluating the Gaussian integral in the above equation.
Let us assume that the local parameterization of $\Sigma$ near $\bar{\bm{q}}$ in terms of the CV is $\psi: \mathbb{R}^{n-m} \to \mathbb{R}^n$, which means that $\bar{\bm{q}} = \psi(\xi)$.
Now, the zero modes at $\bar{\bm{q}}$ belong to the tangent space $T_{\bar{\bm{q}}}\Sigma = \ker \mathsf{C}$ by definition~\cite{leimkuhler2005}.
This implies that the normal modes of the dynamical matrix, which are orthogonal to the zero modes, belong to the $m$-dimensional normal space $N_{\bar{\bm{q}}}\Sigma = (\ker\mathsf{C})^\perp$.
Since $T_{\bar{\bm{q}}}\Sigma \oplus N_{\bar{\bm{q}}}\Sigma = \mathbb{R}^n$, we can always write $\bm{q} = \bm{u} + \bm{v}$ with
$\bm{\bm{u}} \in T_{\bar{\bm{q}}}\Sigma$ and $\bm{v} \in N_{\bar{\bm{q}}}\Sigma$, for any $\bm{q} \in \mathbb{R}^n$ (see Fig.~\ref{sm:fig:levelsets}).
As an orthonormal basis for $N_{\bar{\bm{q}}}\Sigma$, choose the normal modes $\bm{b}_1, \bm{b}_2, \ldots, \bm{b}_m \in \mathbb{R}^n$ and as the basis for $T_{\bar{\bm{q}}}\Sigma$ choose the columns of the $n\times(n-m)$ Jacobian matrix $\nabla\psi(\xi)$.%
\footnote{The normal modes need not be orthonormal if there is degeneracy in the normal frequencies.
But in such a situation, one can still pick a set of orthonormal vectors within the subspace spanned by degenerate normal modes.
With some extra steps, the derivation can also be made to work for an arbitrary basis of $N_{\bar{\bm{q}}}\Sigma$ as well.}
Therefore, we can write
\begin{equation}
  \bm{q} = \bm{u} + \bm{v} = (\nabla\psi)\bm{x} + \mathsf{B}\bm{y},
  \label{sm:eq:coord_change}
\end{equation}
where $\mathsf{B}  = \inmat{\bm{b}_1 & \bm{b}_2 & \cdots & \bm{b}_m}$ is the change-of-basis matrix for $N_{\bar{\bm{q}}}\Sigma$.
The vectors $\bm{x} \in \mathbb{R}^{n-m}$ and $\bm{y} \in \mathbb{R}^m$ represent the components of $\bm{u}$ and $\bm{v}$ in the chosen bases.
Eq.~\eqref{sm:eq:coord_change} suggests the coordinate change $\bm{q} \to (\bm{x}, \bm{y})$.
The quadratic form in the integral of Eq.~\eqref{sm:eq:prob_gaussian} after such a coordinate change can be decomposed into blocks as
\begin{equation}
  \frac{1}{2}
  \begin{pmatrix}
    \bm{x}\trans & \bm{y}\trans
  \end{pmatrix}
  \begin{pmatrix}
    (\nabla\psi)\trans(\nabla\hat{\xi})\trans\nabla{\hat{\xi}}\nabla\psi &
    \quad(\nabla\psi)\trans(\nabla\hat{\xi})\trans\nabla\hat{\xi}\mathsf{B} \\
    \mathsf{B}\trans(\nabla\hat{\xi})\trans\nabla{\hat{\xi}}\nabla\psi &
    \quad\mathsf{B}\trans(\nabla\hat{\xi})\trans\nabla\hat{\xi}\mathsf{B} + \alpha^{-1}\beta\mathsf{D}^\perp &
  \end{pmatrix}
  \begin{pmatrix}
    \bm{x}\\
    \bm{y}
  \end{pmatrix},
  \label{sm:eq:quadratic_form}
\end{equation}
where $\mathsf{D}^\perp$ is the diagonal matrix of the $m$ nonzero eigenvalues of $\mathsf{D}$.
Our assumption is that $\psi(\xi)$ is a valid parameterization of $\Sigma$ that is compatible with the CV map $\hat{\xi}$, i.e., $\bar{\bm{q}} = \psi(\xi)$ and $\hat{\xi}(\bar{\bm{q}}) = \xi$, which means that
\begin{equation}
  (\hat{\xi}\circ\psi)(\xi) = \xi.
\end{equation}
Taking derivatives with respect to $\xi$ on both sides of the above equation we see that
\begin{equation}
  \nabla\hat{\xi}(\bar{\bm{q}})\nabla\psi(\xi) = \mathsf{I}_{n-m},
\end{equation}
where $\mathsf{I}_{n-m}$ is the $(n-m)\times(n-m)$ identity matrix.
Using this, the $n\times n$ block matrix in Eq.~\eqref{sm:eq:quadratic_form} can be written as
\begin{equation}
  \begin{pmatrix}
    \mathsf{I}_{n-m} &
    \quad\nabla\hat{\xi}\mathsf{B}\\
    \mathsf{B}\trans(\nabla\hat{\xi})\trans &
    \quad\mathsf{B}\trans(\nabla\hat{\xi})\trans\nabla\hat{\xi}\mathsf{B} + \alpha^{-1}\beta\mathsf{D}^\perp
  \end{pmatrix}.
\end{equation}
Since $\mathsf{I}_{n-m}$ is trivially invertible, the determinant of the above matrix is
\begin{equation}
  \det\,\mathsf{I}_{n-m}\det\,\big[\mathsf{B} \trans(\nabla\hat{\xi})\trans\nabla\hat{\xi}\mathsf{B}  + \alpha^{-1}\beta\mathsf{D}^\perp - \mathsf{B}\trans(\nabla\hat{\xi})\trans\mathsf{I}^{-1}_{n-m}\nabla\hat{\xi}\mathsf{B}\big]
  =
  \alpha^{-m}\beta^m \det\,\mathsf{D}^\perp.
  \label{sm:eq:determinant}
\end{equation}

Under the coordinate change $\bm{q} \to (\bm{x},\bm{y})$, the volume element $\dd{\bm{q}}$ in the integral in Eq.~\eqref{sm:eq:prob_gaussian} acquires the factor $\sqrt{\abs{\det\,\mathsf{J}\trans\mathsf{J}}}$, where $\mathsf{J}$ is the Jacobian of the transformation.
The matrix $\mathsf{J}\trans\mathsf{J}$ can be readily cast into blocks as
\begin{equation}
  \mathsf{J}\trans\mathsf{J} =
  \begin{pmatrix}
    (\nabla\psi)\trans\nabla\psi & 0 \\
    0 & \mathsf{B}\trans\mathsf{B}
  \end{pmatrix}
  =
  \begin{pmatrix}
    (\nabla\psi)\trans\nabla\psi & 0 \\
    0 & \mathsf{I}_{m}
  \end{pmatrix}.
\end{equation}
Here $(\nabla\psi)\trans\nabla\psi$ is the metric induced by the embedding $\xi \mapsto \psi(\xi)$ and since the normal modes are orthonormal vectors, $\mathsf{B} \trans\mathsf{B}  = \mathsf{I}_m$.
Also, the off-diagonal blocks vanish since they involve inner products of the basis vectors of $T_{\bar{\bm{q}}}\Sigma$ and $N_{\bar{\bm{q}}}\Sigma$, which are orthogonal complements of each other.
This gives $\sqrt{\abs{\det\,\mathsf{J}\trans\mathsf{J}}} = \sqrt{\abs{\det\, (\nabla\psi)\trans\nabla\psi}}$, which together with Eq.~\eqref{sm:eq:determinant} lets us evaluate the Gaussian integral in Eq.~\eqref{sm:eq:prob_gaussian} and write
\begin{equation}
  \mathcal{P}_{\hat{\xi}}(\xi) \sim I(\xi)\left(\frac{2\pi}{\beta}\right)^{m/2}
  \left|\frac{\det\,[\nabla\psi(\xi)]\trans\nabla\psi(\xi)}{\det\,\mathsf{D}^{\perp}(\xi)}\right|^{1/2},
\end{equation}
which completes the derivation of Eq.~\eqref{eq:mpd_regular}.
The form of the above equation suggests that the nonzero eigenvalues of $\mathsf{D}$ can be naturally interpreted as being inversely proportional to the effective widths of the fluctuations along the $m$ dimensions perpendicular to $\Sigma$.
\begin{figure}
  \begin{center}
    \includegraphics{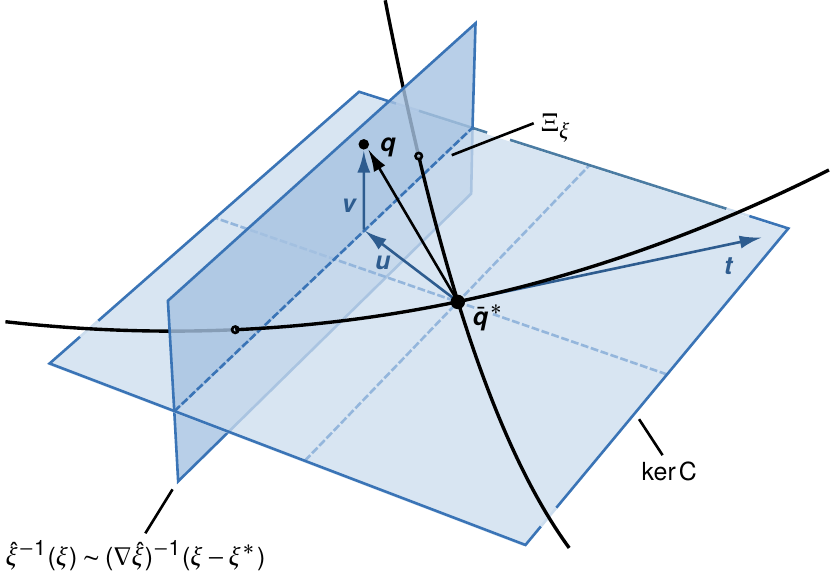}
  \end{center}
  \caption{Cartoon illustrating the geometry of the branches and the level set $(\nabla\hat{\xi})^{-1}(\xi - \xi^{*})$ of the linearized CV map near the singularity, which is at $\bar{\bm{q}}^{*}$.
    The two ground-state configurations in the CV level set are indicated by the intersection points of the branches with $(\nabla\hat{\xi})^{-1}(\xi - \xi^{*})$.  As $\xi \to \xi^{*}$, these ground states get infinitesimally close to each other and the mechanism becomes soft.
    The vector $\bm{u} \in \ker\mathsf{C}$ and the vector $\bm{v} \in (\ker\mathsf{C})^{\perp}$.
    Also indicated is a tangent vector $\bm{t}$ to a branch at the singularity.
    Here there is only one state of self stress and the delta function restricts the integral in Eq.~\eqref{sm:eq:mpd_singular} to the line $\Xi_{\xi}$ formed by the intersection of $(\nabla\hat{\xi})^{-1}(\xi - \xi^{*})$ and $\ker\mathsf{C}$.
  }
  \label{sm:fig:intersect}
\end{figure}

\subsection{Singular values}
\label{sm:sec:singular}

Equation~\eqref{eq:mpd_regular} breaks down as we approach a singularity along $\Sigma$ since the lowest $s$ nonzero eigenvalues of $\mathsf{D}$ become very small and tend to zero, and we need an alternative method to find the marginal densities.
At a singular point of $\Sigma$ with $s$ self stresses, the number of zero modes increases by $s$, which means that now there are only $m-s$ normal modes.
A cartoon of the situation is depicted in Fig.~\ref{sm:fig:intersect}, which shows two branches of a one-dimensional shape space $\Sigma$ intersecting at a singularity $\bar{\bm{q}}^{*}$, where the CV has the value $\xi^{*}$.
Because of the additional self stress, the subspace of zero modes $\ker\mathsf{C}$ is now a plane, which is tangent to both the branches at the singularity.

How should one proceed in such a situation?
At first glance, one might think that all that is required is to replace the harmonic energy in the previous derivation with the quartic energy expansion [Eq.~\eqref{eq:energy_singular}].
This cannot be correct as Eq.~\eqref{eq:energy_singular} is only valid when the expansion is around the singularity, and our goal here is to find $\mathcal{P}_{\hat{\xi}}(\xi)$ not just for the singular CV value $\xi^{*}$, but for all $\xi \to \xi^{*}$.
When $\xi$ is close to $\xi^{*}$, the ground-state configurations in $\hat{\xi}^{-1}(\xi)$ are all regular points of $\Sigma$ that do not have any singular zero modes.
However, at these configurations, the mechanism also becomes very soft in certain directions due to the presence of the nearby singularity, causing the harmonic approximation to break down.
A possible strategy could then be to expand the energy to both harmonic and quartic order along these special soft modes, i.e., those that will ultimately become a zero mode at the singularity as we move along $\Sigma$.
Apart from the complexity of such an expansion, this raises another issue: how should one perform Laplace asymptotics around the ground states in $\hat{\xi}^{-1}(\xi)$?
When $\xi$ is close to $\xi^{*}$, the ground states in $\hat{\xi}^{-1}(\xi)$ get infinitesimally close to each other, as do the branches (see Fig.~\ref{sm:fig:intersect}).
Hence, on doing Laplace asymptotics around each of these ground states and on extending the integration domain to infinity, we would be adding contributions to the marginal density from the same region more than once.
In this sense, the contributions from the branches are not separable in the vicinity of the singularity.

From the above discussion, it should be evident that the approach we used in deriving the harmonic marginal density [Eq.~\eqref{eq:mpd_regular}] will not work for finding an asymptotic expression for $\mathcal{P}_{\hat{\xi}}(\xi)$ for $\xi \to \xi^{*}$.
The only possibility then is to rely on Eq.~\eqref{sm:eq:coarea}, which expresses $\mathcal{P}_{\hat{\xi}}(\xi)$ as an exact integral over the CV level set $\hat{\xi}^{-1}(\xi)$.
Note again that Eq.~\eqref{sm:eq:coarea} makes no reference to the shape space $\Sigma$ as such, enabling us to side step the issues described above.
This however, comes at the expense of having to do a higher-dimensional integral over $\hat{\xi}^{-1}(\xi)$.
The next strategy is to make physically valid assumptions that can be used to dimensionally reduce this integral.

The first such assumption made in the main text was that the CV map can be approximated by its Taylor expansion $\hat{\xi} = \xi^{*} + (\nabla\hat{\xi})\bm{q} + \mathcal{O}(\Abs{\bm{q}}^2)$ around the singularity $\bar{\bm{q}}^{*}$ for points close to $\bar{\bm{q}}^{*}$.
This linearizes the CV map and turns its level sets $(\nabla\hat{\xi})^{-1}(\xi - \xi^{*}) = \{\bm{u} \in \mathbb{R}^{n}: \xi^{*} + (\nabla\hat{\xi})\bm{u} = \xi\}$ near the singularity into hyperplanes (see Fig.~\ref{sm:fig:intersect}).
Since this hyperplane is considerably easier to parameterize in comparison to the actual CV level set $\hat{\xi}^{-1}(\xi)$, this makes the evaluation of $\mathcal{P}_{\hat{\xi}}(\xi)$ as a surface integral using Eq.~\eqref{sm:eq:coarea} less difficult.
The second assumption is that the fast modes $\bm{v} \in (\ker\mathsf{C})^{\perp}$ are such that they do not change the value of the CV to linear order at the singularity, i.e., $(\nabla\hat{\xi})\bm{v} = \bm{0}$.
This enables us to write $\bm{q} = \bm{u} + \bm{v}$, with $\bm{u} \in \ker\mathsf{C}$ and $\bm{v} \in (\ker\mathsf{C})^\perp$, and use the lowest-order approximation of the energy near the singularity [Eq.~\eqref{eq:energy_singular}] to get\footnote{The Jacobian factor introduced by the transformation $\bm{q} \to (\bm{u}, \bm{v})$ is unity if we pick orthonormal bases for $\ker\mathsf{C}$ and $(\ker\mathsf{C})^{\perp}$.}
\begin{equation}
  \hspace{-0.5em}
  \begin{aligned}
    \mathcal{P}_{\hat{\xi}}(\xi) &= \int \dd\bm{q}\, I(\bm{q})\,\delta\left[\hat{\xi}(\bm{q}) - \xi\right] \exp\left[-\beta U(\bm{q})\right]\\
                                 &\sim I(\xi^{*}) \int_{\ker\mathsf{C}} \dd\bm{u}\, \delta\left[(\nabla\hat{\xi})\bm{u} - (\xi - \xi^{*})\right] \int_{(\ker\mathsf{C})^{\perp}} \dd\bm{v}\,  \exp\left\{-\frac{1}{2}\beta\kappa[\mathsf{C}\bm{v} + \bm{w}(\bm{u})]\trans[\mathsf{C}\bm{v} + \bm{w}(\bm{u})]\right\},
  \end{aligned}
\end{equation}
where we have used $(\nabla\hat{\xi})(\bm{u} + \bm{v}) = (\nabla\hat{\xi})\bm{u}$.
As we can see from the above equation, the second assumption has allowed us to separate the contributions to $\mathcal{P}_{\hat{\xi}}(\xi)$ from the fast vibrational modes in $(\ker\mathsf{C})^{\perp}$ and the zero modes in $\ker\mathsf{C}$.
As before, we choose the $m - s$ normal modes as the basis for $(\ker\mathsf{C})^{\perp}$ and write $\bm{v} = \mathsf{B}\bm{y}$, with $\mathsf{B}$ being the change-of-basis matrix and $\bm{y} \in \mathbb{R}^{m-s}$ representing the components of $\bm{v}$ in the chosen basis.
This turns the integral over $(\ker\mathsf{C})^{\perp}$ into an $(m-s)$-dimensional Gaussian integral over $\bm{y}$, which after a straightforward integration yields
\begin{equation}
  \hspace{-0.5em}
  \begin{aligned}
    \mathcal{P}_{\hat{\xi}}(\xi) &\sim I(\xi^{*}) \int_{\ker\mathsf{C}} \dd\bm{u}\, \delta\left[(\nabla\hat{\xi})\bm{u} - (\xi - \xi^{*})\right] \int_{\mathbb{R}^{m-s}} \dd\bm{y}\, \exp\left\{-\frac{1}{2}\beta\kappa[\mathsf{C}\mathsf{B}\bm{y} + \bm{w}(\bm{u})]\trans[\mathsf{C}\mathsf{B}\bm{y} + \bm{w}(\bm{u})]\right\}\\
                                             &= \frac{I(\xi^{*})}{\left|\det\, \mathsf{D}^\perp\right|^{1/2}} \left(\frac{2\pi}{\beta}\right)^{(m-s)/2} \int_{\ker\mathsf{C}} \dd\bm{u}\, \delta\left[(\nabla\hat{\xi})\bm{u} - (\xi - \xi^{*})\right] \exp\left[-\frac{1}{2}\beta\kappa \bm{w}(\bm{u})\trans\Pi \bm{w}(\bm{u})\right],
  \end{aligned}
  \label{sm:eq:mpd_gaussian}
\end{equation}
where $\mathsf{D}^\perp$ is the diagonal matrix of the $m - s$ nonzero eigenvalues of $\mathsf{D}$ at the singularity and
\begin{equation}
  \Pi = \mathsf{I}_{m} - \mathsf{C}\mathsf{B} (\mathsf{B} ^\mathsf{T}\mathsf{C}^\mathsf{T}\mathsf{C}\mathsf{B} )^{-1}\mathsf{B} ^\mathsf{T}\mathsf{C}^\mathsf{T}
\end{equation}
is a projection operator that projects $\bm{w}(\bm{u})$ to the cokernel of $\mathsf{C}\mathsf{B}$ (i.e., to $\ker\mathsf{B}\trans\mathsf{C}\trans$).
Note that this operator would trivially have been the identity matrix if it were not for the cross term $\bm{w}(\bm{u})\trans\mathsf{K}\mathsf{C}\bm{v}$ in the energy expansion [Eq.~\eqref{eq:energy_singular}] that couples the fast modes and the zero modes.
Since the cokernel of $\mathsf{CB}$ is identical to the cokernel of $\mathsf{C}$, it is spanned by an orthonormal basis of self stresses $\bm{\sigma}$, and one can write the action of $\Pi$ on $\bm{w}(\bm{u})$ as
\begin{equation}
  \Pi \bm{w}(\bm{u)} = \sum_{\bm{\sigma}} \bm{\sigma}[\bm{\sigma}\cdot\bm{w}(\bm{u})].
  \label{sm:eq:mpd_proj}
\end{equation}
Using this in Eq.~\eqref{sm:eq:mpd_gaussian} and after a straightforward application of the coarea formula [Eq.~\eqref{sm:eq:coarea}] we find to the lowest order, for $\xi \to \xi^{*}$,
\begin{equation}
  \begin{aligned}
    \mathcal{P}_{\hat{\xi}}(\xi) &\sim \frac{I(\xi^{*})}{\left|\det\, \mathsf{D}^\perp\right|^{1/2}} \left(\frac{2\pi}{\beta}\right)^{(m-s)/2} \int_{\ker\mathsf{C}} \dd\bm{u}\, \delta\left[(\nabla\hat{\xi})\bm{u} - (\xi - \xi^{*})\right] \exp\left\{-\frac{1}{2}\beta\kappa\sum_{\bm{\sigma}}[\bm{\sigma}\cdot \bm{w}(\bm{u})]^2\right\}\\
                                 &= \frac{I({\xi}^{*})}{\left|\det\, \mathsf{D}^\perp \det\, \nabla\hat{\xi}(\nabla\hat{\xi})\trans\right|^{1/2}}  \left(\frac{2\pi}{\beta}\right)^{(m-s)/2} \int_{\Xi_{\xi}} \dd\Omega(\bm{u})\, \exp\left\{-\frac{1}{2}\beta\kappa\sum_{\bm{\sigma}}[\bm{\sigma}\cdot\bm{w}(\bm{u})]^2\right\}.
  \end{aligned}
  \label{sm:eq:mpd_singular}
\end{equation}
Above, the integration domain $\Xi_{\xi} = (\nabla\hat{\xi})^{-1}(\xi - \xi^{*}) \cap \ker\mathsf{C}$, is a hyperplane of $(n - m + s) - (n - m) = s$ dimensions and $\dd\Omega(\bm{u})$ is the surface measure on $\Xi_{\xi}$.
After choosing a convenient parameterization for $\Xi_{\xi}$ in terms of the components of $\bm{u}$ in some basis of $\ker\mathsf{C}$, Eq.~\eqref{sm:eq:mpd_singular} becomes an $s$-dimensional integral involving the exponential of a quartic polynomial, which should converge if $\Sigma_{\bm{\sigma}} [\bm{\sigma}\cdot\bm{w}(\bm{u)}]^2$ does not identically vanish in some region of $\Xi_{\xi}$ that extends to infinity (see below for an extended discussion on this).
This completes the derivation of Eq.~\eqref{eq:mpd_singular}.
In deriving Eq.~\eqref{eq:mpd_singular}, unlike in the derivation of Eq.~\eqref{eq:mpd_regular}, we have not looked at contributions to $\mathcal{P}_{\hat{\xi}}(\xi)$ from the ground states on each branch of $\Sigma$ individually.
Hence, Eq.~\eqref{eq:mpd_singular} represents the collective contribution to the marginal density from all the branches that intersect to form the singularity at $\bar{\bm{q}}^{*}$.

\subsubsection{Convergence of the integral in \texorpdfstring{Eq.~\eqref{eq:mpd_singular}}{Eq. (5)}}
\label{sm:sec:convergence}

Before discussing the conditions that are required for the integral in Eq.~\eqref{eq:mpd_singular} to converge, we digress slightly and discuss second-order rigidity, a frequently invoked notion in rigidity theory~\cite{connelly1994,connelly1996}.
In the notation that we have been using, a bar-joint framework is considered to be second-order rigid if there are no vector pairs $(\bm{u}, \bm{v})$ with $\bm{u} \in \ker\mathsf{C}$ and $\bm{v} \in (\ker\mathsf{C})^{\perp}$ such that\footnote{Note that this equation is nothing but the Taylor expansion of the constraint map $f$ to the lowest order in $\bm{u}$ and $\bm{v}$.}
\begin{equation}
  \mathsf{C}\bm{v} + \bm{w}(\bm{u}) = \bm{0}.
\end{equation}
For a singular configuration of the framework that supports nonzero self stresses $\bm{\sigma}$, second-order rigidity is equivalent to saying that there is no zero mode $\bm{u} \in \ker\mathsf{C}$ such that\footnote{See, e.g., Corollary 5.2.2 of Ref.~\cite{connelly1996} or Corollaries 4.15--4.17 of Ref.~\cite{williams2003}.}
\begin{equation}
  \bm{\sigma}\cdot\bm{w}(\bm{u}) = 0,\label{sm:eq:2ndorder}
\end{equation}
for all $\bm{\sigma} \in \ker\mathsf{C}\trans$.
A well-known result is that a bar-joint framework is rigid to all orders if it is second-order rigid~\cite{connelly1996}.

Clearly, a mechanism is not rigid (to any order) by definition and all tangents to the branches of the shape space $\Sigma$ at a singularity $\bar{\bm{q}}^{*}$ satisfy Eq.~\eqref{sm:eq:2ndorder}.
Even though one can speak of tangent vectors to the branches of $\Sigma$ at $\bar{\bm{q}}^{*}$, the shape space $\Sigma$ itself ceases to be a smooth manifold at $\bar{\bm{q}}^{*}$.
Hence, there is no well-defined tangent space at $\bar{\bm{q}}^{*}$ and it is common practice to consider instead the solution space of Eq.~\eqref{sm:eq:2ndorder},
\begin{equation}
  \mathcal{T} = \left\{\bm{t} \in \ker\mathsf{C} : \bm{\sigma}\cdot \bm{w}(\bm{t}) = 0 \text{ for all } \bm{\sigma} \in \ker\mathsf{C}\trans \right\},
  \label{sm:eq:tangentcone}
\end{equation}
which is the set of zero modes that preserve the constraints to second-order~\cite{chen2018}.
For this reason, $\mathcal{T}$ is often called the second-order tangent cone.\footnote{See, e.g., Section 4.2 of Ref.~\cite{wu2020}; also see the related discussions in Refs.~\cite{muller2017,muller2019,lopez-custodio2020}.
Tangent cones themselves were originally introduced by \citet{whitney1965} to study tangents to analytic varieties.}
Even though the tangents to the branches of $\Sigma$ belong to $\mathcal{T}$, in general, there could be vectors in $\mathcal{T}$ that are not tangents.
In such situations, the tangents cannot be resolved by considering the second-order tangent cone alone and higher-order analysis is necessary~\cite{muller2017,lopez-custodio2020}.
To simplify things, from here on we assume that the tangents can be resolved at second order, i.e., every element of $\mathcal{T}$ is a tangent to the branches of the shape space $\Sigma$ at the singularity.
For instance, in the cartoon in Fig.~\ref{sm:fig:intersect}, $\mathcal{T}$ would be the union of the two tangents at the singularity.

Now, the integral that defines $\mathcal{P}_{\hat{\xi}}(\xi)$ also includes a delta function $\delta[\hat{\xi}(\bm{q}) - \xi]$.
After linearizing the CV map $\hat{\xi}$ around the singularity and integrating out the fast modes, the domain of integration in Eq.~\eqref{eq:mpd_singular} becomes the hyperplane $\Xi_{\xi} = \ker\mathsf{C} \cap (\nabla\hat{\xi})^{-1}(\xi - \xi^{*})$.
At a singularity with $s$ self stresses $\bm{\sigma}_{1}, \bm{\sigma}_{2}, \ldots, \bm{\sigma}_{s}$, consider the map $g: \ker\mathsf{C} \to \mathbb{R}^{n - m + s}$ defined by $g(\bm{u}) = [(\nabla\hat{\xi})\bm{u} - (\xi - \xi^{*}), \bm{\sigma}_{1}\cdot \bm{w}(\bm{u}), \bm{\sigma}_{2}\cdot \bm{w}(\bm{u}), \ldots, \bm{\sigma}_{s}\cdot \bm{w}(\bm{u})]$.
In a basis of $\ker\mathsf{C}$, the equation $g(\bm{u}) = \bm{0}$ defines an exactly determined system of equations in $n-m+s$ unknowns, whose solutions (if they exist) are tangents that belong to $\Xi_{\xi}$.
If this equation has isolated roots, then the term $\sum_{\bm{\sigma}} [\bm{\sigma}\cdot\bm{w}(\bm{u})]^{2}$ in the exponential of Eq.~\eqref{eq:mpd_singular} is zero only for a discrete set of tangent vectors in $\Xi_{\xi}$ and is positive everywhere else, making the integral convergent.
(In Fig.~\ref{sm:fig:intersect}, this discrete set is composed of the tangents to the two branches, which when extended intersect the line $\Xi_{\xi}$, which is the domain of integration.)
A necessary condition for the equation $g(\bm{u}) = \bm{0}$ to have isolated roots in $\Xi_{\xi}$ is that $(\nabla\hat{\xi})\bm{t} \neq \bm{0}$ for all $\bm{t} \in \mathcal{T}$.
Given how $g(\bm{u})$ is a system of $n - m$ linear and $s$ quadratic equations, a stronger general condition that ensures this eludes us at present.
On the other hand, if there is a tangent $\bm{t}$ such that $(\nabla\hat{\xi})\bm{t} = \bm{0}$, the integral diverges.
Such cases are pathological and indicate a poor CV choice.
After all, if $\bm{t}$ is a tangent to $\Sigma$, then it is a slow mode that corresponds to a shape change in the mechanism, and one would definitely want the value of the CV to change along it.

To conclude, the convergence of Eq.~\eqref{eq:mpd_singular} relies on the term $\sum_{\bm{\sigma}}[\bm{\sigma}\cdot\bm{w}(\bm{u})]^{2}$ in the exponential vanishing only for a finite number of isolated points in the integration domain $\Xi_{\xi}$.
Two necessary conditions required for this are: (i) tangents to the branches of the shape space at the singularity can be resolved at second order and form the solution space $\mathcal{T}$ of Eq.~\eqref{sm:eq:2ndorder}, and (ii) the CV map is such that $(\nabla\hat{\xi})\bm{t} \neq 0$ for all $\bm{t} \in \mathcal{T}$.

\subsubsection{Scaling of the marginal density}
\label{sm:sec:scaling}

To see how the marginal density $\mathcal{P}_{\hat{\xi}}(\xi^{*})$ at a singular value $\xi^{*}$ scales with $\beta$ and $\kappa$, we first choose a basis for $\ker\mathsf{C}$ so that $\bm{u} = \mathsf{A}\bm{x}$, where $\bm{x} \in \mathbb{R}^{n-m+s}$ represents the components of $\bm{u}$ in the chosen basis and $\mathsf{A}$ is the associated change-of-basis matrix.
Now, the $s$-dimensional hyperplane $\Xi_{\xi}$ formed by the intersection of the linearized CV level set and $\ker\mathsf{C}$ is defined by $\nabla\hat{\xi}\mathsf{A}\bm{x} = \bm{0}$, which is a set of $n-m$ homogeneous linear equations in $n-m+s$ variables.
Without loss of generality, let us assume that we can solve these equations to obtain the last $n - m$ components of $\bm{x}$ in terms of its first $s$ components $\tilde{\bm{x}} = (x_{1}, x_{2}, \ldots, x_{s})$.
This enables us to parameterize the hyperplane $\Xi_{\xi}$ using $\tilde{\bm{x}}$.
As each component of the vector $\bm{w}(\bm{u})$ is a quadratic form in $\bm{x}$, after the elimination step, the term $\sum_{\bm{\sigma}}[\bm{\sigma}\cdot\bm{w}(\bm{u})]^{2}$ in the exponential of Eq.~\eqref{eq:mpd_singular} becomes a homogeneous quartic polynomial $\widetilde{U}(\tilde{\bm{x}})$.
A rescaling of the components $\tilde{\bm{x}} \to (\beta\kappa)^{-1/4}\tilde{\bm{x}}$, changes the surface measure from $\dd\Omega(\tilde{\bm{x}}) \to (\beta\kappa)^{-s/4}\dd\Omega(\tilde{\bm{x}})$ and turns $\widetilde{U}(\tilde{\bm{x}}) \to (\beta\kappa)^{-1}\widetilde{U}(\tilde{\bm{x}})$, yielding
\begin{equation}
  \mathcal{P}_{\hat{\xi}}(\xi^{*}) \sim \frac{I({\xi}^{*})}{\left|\kappa^{s/2}\det\,\mathsf{D}^\perp \det\, \nabla\hat{\xi}(\nabla\hat{\xi})\trans\right|^{1/2}}  \left(\frac{2\pi}{\beta}\right)^{m/2 - s/4} \int_{\Xi_{\xi}} \dd\Omega(\tilde{\bm{x}})\, \exp\left[-\frac{1}{2}\widetilde{U}(\tilde{\bm{x}})\right].
  \label{sm:eq:scaling}
\end{equation}
Clearly, the above integral is purely geometric in nature and all $\beta$ dependence has been extracted
Finally, noting that $\det\,\mathsf{D}^{\perp} \sim \kappa^{m - s}$ we see that the marginal density $\mathcal{P}_{\hat{\xi}}(\xi^{*}) \sim (\beta\kappa)^{-m/2 + s/4}$.
It should be emphasized that we can only do this analysis for $\xi = \xi^{*}$.
For other values of $\xi$ close to $\xi^{*}$, the hyperplane $\Xi_{\xi}$ is defined by the inhomogeneous equation $\nabla\hat{\xi}\mathsf{A}\bm{x} = \xi - \xi^{*}$, which makes $\widetilde{U}(\tilde{\bm{x}})$ a similarly inhomogeneous quartic polynomial, making a rescaling argument impossible.

The marginal density for regular values of the CV, which scales like $\mathcal{P}_{\hat{\xi}}(\xi) \sim (\beta\kappa)^{-m/2}$ [Eq.~\eqref{eq:mpd_regular}], is always subdominant to the marginal density at a singular value $\xi^{*}$, which scales like $\mathcal{P}_{\hat{\xi}}(\xi^{*}) \sim (\beta\kappa)^{-m/2 + s/4}$ for $s > 0$.
Hence, we see that the softening of the mechanism at a singularity causes an energetic free-energy barrier to develop between regular and singular values of the CV, with a temperature/stiffness dependence $\sim \ln \beta\kappa$.
In comparison, the free-energy barriers between singular values of the CV are independent of $\beta$ and $\kappa$, and depend only on geometric parameters if the corresponding configurations have the same number of self stresses $s$.
Although this might lead us to conclude that such barriers are entropic in origin, note that there would be energetic barriers along most realizable transition paths separating these configurations.

The stark difference in the asymptotic scaling of $\mathcal{P}_{\hat{\xi}}(\xi)$ at a singular value $\xi = \xi^{*}$ and for values farther from it shows that the true scaling and behavior of $\mathcal{P}_{\hat{\xi}}(\xi)$ for intermediate values of $\xi$ is nontrivial.
A natural question is then: for what values of $\xi$ would the harmonic and quartic approximations capture the true behavior of $\mathcal{P}_{\hat{\xi}}(\xi)$?
Equation~\eqref{eq:mpd_regular}, derived using the harmonic approximation and a direct application of Laplace's method, is only accurate so long as the lowest nonzero eigenvalue $\omega_{\text{min}}(\xi)$ of the dynamical matrix $\mathsf{D}$ is such that $\beta\omega_{\text{min}}(\xi)$ is very large.
This is also what causes it to break down as we approach a singularity, near which $\omega_{\text{min}}(\xi)$ monotonically\footnote{For one-dimensional shape spaces, using Rayleigh--Schr\"{o}dinger perturbation theory~\cite{dirac1958} and considering the dynamical matrix at the singularity as the ``unperturbed Hamiltonian'', it can be shown that $\omega_{\text{min}} \sim (\xi - \xi^{*})^{2}$ for $\xi \to \xi^{*}$.} decreases to zero as $\xi \to \xi^{*}$.
This also implies that as $\beta$ becomes larger, the harmonic approximation starts capturing the true behavior of the marginal density for a larger range of $\xi$ values.
For very large $\beta$, the range of validity of the quartic approximation is also bound to increase as the errors in the approximation become small.
This means that, for large $\beta$, we expect to see some amount of overlap in the marginal density estimates using Eqs.~\eqref{eq:mpd_regular} and \eqref{eq:mpd_singular} as evidenced by the free-energy curves in Figs.~\ref{fig:4bar_free} and \ref{fig:origami}(c).
We expect the exact nature of the overlap to be problem specific with a strong dependence on $\beta$ and we leave a more thorough analysis for future work.

\section{Planar Four-Bar Linkage}
\label{sm:sec:4bar}

\subsection{Body frame}

\begin{figure}
  \begin{center}
    \includegraphics{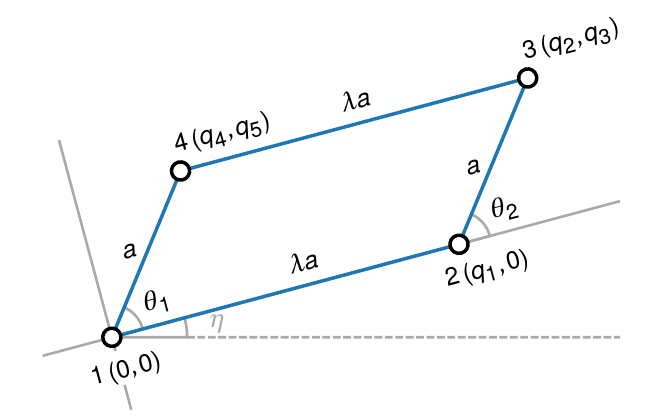}
  \end{center}
  \caption{Body frame on the planar four-bar linkage with joint~1 at the origin and bar~1--2 lying along the horizontal axis.
    The angle between the horizontal axes of the body and lab frame is $\eta$.}
  \label{sm:fig:4bar}
\end{figure}

Rigid motions can be integrated out by transforming to a local Cartesian coordinate system (body frame) attached to the four-bar linkage with joint~1 at the origin and bar~1--2 lying along the horizontal axis as shown in Fig.~\ref{sm:fig:4bar}.
Let $(r_{i1}, r_{i2}),\, i = 1, 2, 3, 4$ be the coordinates of the four joints in the lab frame.
The configuration vector $\bm{q} \in \mathbb{R}^{5}$ of the linkage in the body frame is $\bm{q} = (q_{1}, q_{2}, \ldots, q_{5})$.
Also, two translational coordinates $x_{1}, x_{2}$ specify the position of joint~1, and an orientational coordinate $\eta$, which is the angle between the horizontal axes of the lab and body frames, gives the overall rotation of the linkage.
The explicit coordinate transformation $\bm{r} \to (x_{1}, x_{2}, \eta, \bm{q})$ is given by
\begin{equation}
  \begin{aligned}
    \inmat{r_{11} \\ r_{12}} &= \inmat{x_{1} \\ x_{2}}, {}&
    \inmat{r_{21} \\ r_{22}} &= \inmat{x_{1} \\ x_{2}} + \mathsf{R}(\eta)\inmat{q_1 \\ 0},\\
    \inmat{r_{31} \\ r_{32}} &= \inmat{x_{1} \\ x_{2}} + \mathsf{R}(\eta)\inmat{q_2 \\ q_3}, {}&
    \inmat{r_{41} \\ r_{42}} &= \inmat{x_{1} \\ x_{2}} + \mathsf{R}(\eta)\inmat{q_4 \\ q_5}.
  \end{aligned}
  \label{sm:eq:4bar_trans}
\end{equation}
Here $\mathsf{R}(\eta)$ is the rotation matrix in $\mathbb{R}^2$.
Dropping the constant factor that one gets after integrating over $x_1, x_2, \text{and } \eta$, the overall Jacobian factor involved in the transformation given by Eq.~\eqref{sm:eq:4bar_trans} is
\begin{equation}
  I(\bm{q}) = \abs{q_1}.
  \label{sm:eq:4bar_jacobian}
\end{equation}

\subsection{Branch parameterization}

The four-bar linkage admits two modes of deformations, giving its shape space a branched appearance.
On the parallel branch the angles $\theta_{1}$ and $\theta_{2}$ are equal, whereas on the twisted branch they have a nonlinear relationship with opposite signs.
To find the exact relationship between $\theta_{1}$ and $\theta_{2}$ on the twisted branch, we first write down the constraint equation for bar~3--4 (see Fig.~\ref{sm:fig:4bar}) of the four-bar linkage:
\begin{equation}
  [\lambda a + a(\cos{\theta_2} - \cos{\theta_1})]^2 + (a\sin{\theta_2} - a\sin{\theta_1})^2 = \lambda^2 a^2.
\end{equation}
Assuming that the aspect ratio $\lambda \ne 1$, this equation can be simplified and factorized in two different ways to get
\begin{subequations}
\begin{align}
  \left[\cos{\theta_2} - \cos{\theta_1}\right]\left[\cos{\theta_2} - \frac{(1 + \lambda^2)\cos{\theta_1} - 2\lambda}{1 - 2\lambda\cos{\theta_1} + \lambda^2}\right] &= 0,\\
  \left[\sin{\theta_2} - \sin{\theta_1}\right]\left[\sin{\theta_2} - \frac{(1 - \lambda^2)\sin{\theta_1}}{1 -2\lambda\cos{\theta_1} + \lambda^2}\right] &= 0.
\end{align}
\end{subequations}
The solutions to the above equations tell us the relationship between the input angle $\theta_1$ and the output angle $\theta_2$ on the two branches, e.g., on the parallel branch
\begin{equation}
  \cos\theta_2 = \cos\theta_1,\qquad
  \sin\theta_2 = \sin\theta_1\,;
  \label{sm:eq:parallel}
\end{equation}
and on the twisted branch
\begin{equation}
  \cos\theta_2 = \frac{(1 + \lambda^2)\cos{\theta_1} - 2\lambda}{1 - 2\lambda\cos{\theta_1} + \lambda^2}, \qquad
  \sin\theta_2 = \frac{(1 - \lambda^2)\sin{\theta_1}}{1 - 2\lambda\cos{\theta_1} + \lambda^2},
  \label{sm:eq:twisted}
\end{equation}
which is what we intended to find.
We can express a point $\bar{\bm{q}}$ in the shape space $\Sigma \subset \mathbb{R}^5$ of the four-bar linkage in terms of the angles $\theta_1$ and $\theta_2$ as $\bar{\bm{q}} = [\lambda a,\, a(\lambda + \cos{\theta_2}),\, a\sin{\theta_2},\, a\cos{\theta_1},\, a\sin{\theta_1}]$.
Using Eqs.~\eqref{sm:eq:parallel} and \eqref{sm:eq:twisted} we find two parameterizations for $\Sigma$, namely $\psi_+: \mathbb{R} \to \mathbb{R}^5$ (parallel branch) and $\psi_-: \mathbb{R} \to \mathbb{R}^5$ (twisted branch), defined by
\begin{subequations}
  \begin{align}
    \psi_+(\theta_1) &= \left[\lambda a, a(\lambda + \cos{\theta_1}), a\sin{\theta_1}, a\cos{\theta_1}, a\sin{\theta_1}\right],\label{sm:eq:4bar_param_parallel}\\
    \psi_-(\theta_1) &= \left[\lambda a, a\frac{(1-\lambda^2)(\cos{\theta_1} - 1)}{1-2\lambda\cos{\theta_1}+\lambda^2}, a\frac{(1-\lambda^2)\sin{\theta_1}}{1-2\lambda\cos{\theta_1}+\lambda^2}, a\cos{\theta_1}, a\sin{\theta_1}\right].\label{sm:eq:4bar_param_twisted}
  \end{align}
\end{subequations}
The above equations define two curves in $\mathbb{R}^{5}$ parameterized by the angle $\theta_{1}$.
The induced metric on these two curves can be readily computed as
\begin{subequations}
  \begin{align}
    (\nabla\psi_{+})\trans\nabla\psi_+ &= \Abs{\partial{\psi_{+}}/\partial{\theta_{1}}}^{2} = 2 a^2,\label{sm:eq:4bar_ind1}\\
    (\nabla\psi_{-})\trans\nabla\psi_- &= \Abs{\partial{\psi_{-}}/\partial{\theta_{1}}}^{2} = \frac{2a^2[1-2\lambda(1-\lambda\cos{\theta_1}+\lambda^2)\cos{\theta_1} + \lambda^4]}{(1-2\lambda\cos{\theta_1}+ \lambda^2)^2}.\label{sm:eq:4bar_ind2}
  \end{align}
\end{subequations}

\subsection{Marginal probability densities}

\subsubsection{Regular values}

Now that we have explicit parameterizations of the two branches of the four-bar linkage in terms of the CV $\theta_{1}$, we can find the marginal density $\mathcal{P}_{\hat{\theta}_{1}}(\theta_{1})$ at regular values of $\theta_{1}$, far from the singular values, i.e., for $0 \ll \abs{\theta_{1}} \ll \pi$.
Note that for each value of $\theta_{1}$, there are two ground states---one on the parallel branch and one on the twisted branch.
This means that we need to separately find the contributions of these ground states using Eq.~\eqref{eq:mpd_regular} and then add them together to obtain $\mathcal{P}_{\hat{\theta}_{1}}(\theta_{1})$.

Starting with constraint function for bar~1--2 and going around the linkage in counterclockwise order, we find the constraint map $f: \mathbb{R}^{5} \to \mathbb{R}^{4}$ in the body frame to be
\begin{equation}
  f(\bm{q}) = \left[\frac{q_1^2 - \lambda^2 a^2}{2\lambda a},\, \frac{(q_2 - q_1)^2 + q_3^2 - a^2}{2a},\, \frac{(q_4 - q_2)^2 + (q_5 - q_3)^2 - \lambda^2 a^2}{2\lambda a},\, \frac{q_4^2 + q_5^2 - a^2}{2a}\right],
\end{equation}
and the compatibility matrix $\mathsf{C}$ to be
\begin{equation}
  \mathsf{C} = \nabla f = a^{-1}\begin{pmatrix}
    \lambda^{-1}q_1 & 0 & 0 & 0 & 0 \\
    (q_1-q_2) & q_2-q_1 & q_3 & 0 & 0\\
    0 & \lambda^{-1}(q_2-q_4) & \lambda^{-1}(q_3-q_5) & \lambda^{-1}(q_4-q_2) & \lambda^{-1}(q_5-q_3)\\
  0 & 0 & 0 & q_4 & q_5
  \label{sm:eq:4bar_compatibility}
\end{pmatrix}.
\end{equation}
At regular points $\mathsf{C}$ has full rank, which implies that the dynamical matrix $\mathsf{D} = \mathsf{C}\trans\mathsf{K}\mathsf{C}$ and the matrix $\mathsf{K}\mathsf{C}\mathsf{C}\trans$ have the same nonzero eigenvalues.
This gives $\det\,\mathsf{D}^\perp = \det\,\mathsf{K}\mathsf{C}\mathsf{C}\trans = \kappa^4\det\,\mathsf{C}\mathsf{C}\trans$.
Inserting the parameterizations from Eqs.~\eqref{sm:eq:4bar_param_parallel} and \eqref{sm:eq:4bar_param_twisted} into the compatibility matrix we compute $\det\,\mathsf{D}^\perp$ along the two branches as
\begin{subequations}
\begin{align}
  \det\,\mathsf{D}^\perp_+ &= 2\kappa^4\sin^2\theta_1,\\
  \det\,\mathsf{D}^\perp_- &= \frac{2\kappa^4\sin^2{\theta_1}[1-2\lambda(1-\lambda\cos{\theta_1}+\lambda^2)\cos{\theta_1} + \lambda^4]}{(1-2\lambda\cos{\theta_1}+\lambda^2)^2}.
\end{align}
\end{subequations}
The asymptotic marginal density is then
\begin{equation}
  \begin{aligned}
    \mathcal{P}_{\hat{\theta}_1}(\theta_1) &\sim I(\theta_1)\left(\frac{2\pi}{\beta}\right)^2\left[\left|\frac{\det\,(\nabla\psi_{+})\trans\nabla\psi_+}{\det\,\mathsf{D}_+^\perp}\right|^{1/2} + \left|\frac{\det\,(\nabla\psi_{-})\trans\nabla\psi_-}{\det\,\mathsf{D}_-^\perp}\right|^{1/2}\right]\\
                                           &= 2\lambda a^{2}\left(\frac{2\pi}{\beta\kappa}\right)^2|\sin{\theta_1}|^{-1},
  \end{aligned}
  \label{sm:eq:4bar_r}
\end{equation}
where we have used $I(\theta_{1}) = \abs{q_{1}} = \lambda a$ [Eq.~\eqref{sm:eq:4bar_jacobian}] and the expressions for the induced metrics [Eqs.~\eqref{sm:eq:4bar_ind1} and \eqref{sm:eq:4bar_ind2}] computed earlier.

\subsubsection{Singular values}

The four-bar linkage has shape-space singularities at $\theta_1 = 0$ and $\theta_1 = \pm\pi$ corresponding to configurations where the bars are collinear.
Let us first look at the singularity at $\theta_1 = 0$, where the configuration vector $\bar{\bm{q}}^{*} = [\lambda a, a(\lambda + 1), 0, a, 0]$.
The compatibility and dynamical matrices at this point are
\begin{equation}
  \mathsf{C} =
  \begin{pmatrix}
  1  & 0 & 0 & 0  & 0 \\
  -1 & 1 & 0 & 0  & 0 \\
  0  & 1 & 0 & -1 & 0 \\
  0  & 0 & 0 & 1  & 0 \\
  \end{pmatrix}
  \quad\text{and}\quad
  \mathsf{D} = \mathsf{C}\trans\mathsf{K}\mathsf{C} = \kappa
  \begin{pmatrix}
    2  & -1 & 0 & 0  & 0 \\
    -1 & 2  & 0 & -1 & 0 \\
    0  & 0  & 0 & 0  & 0 \\
    0  & -1 & 0 & 2  & 0 \\
    0  & 0  & 0 & 0  & 0 \\
  \end{pmatrix}.
\end{equation}
The dynamical matrix has nonzero eigenvalues $(2+\sqrt{2})\kappa$, $2\kappa$, and $(2-\sqrt{2})\kappa$, which gives $\det\,\mathsf{D}^\perp = 4\kappa^3$.
Also, the Hessian matrices of the four constraint functions $f_i$ at the singularity are
\begin{equation}
\begin{aligned}
  \hess f_1 &=
  (\lambda a)^{-1}\begin{pmatrix}
    1 & 0 & 0 & 0 & 0 \\
    0 & 0 & 0 & 0 & 0 \\
    0 & 0 & 0 & 0 & 0 \\
    0 & 0 & 0 & 0 & 0 \\
    0 & 0 & 0 & 0 & 0 \\
  \end{pmatrix},\qquad {}&
  \hess f_2 &= a^{-1}
  \begin{pmatrix}
    1 & -1 & 0 & 0 & 0 \\
    -1 & 1 & 0 & 0 & 0 \\
    0 & 0 & 1 & 0 & 0 \\
    0 & 0 & 0 & 0 & 0 \\
    0 & 0 & 0 & 0 & 0 \\
  \end{pmatrix},\qquad\\
  \hess f_3 &=
  (\lambda a)^{-1}\begin{pmatrix}
    0 & 0 & 0 & 0 & 0 \\
    0 & 1 & 0 & -1 & 0 \\
    0 & 0 & 1 & 0 & -1 \\
    0 & -1 & 0 & 1 & 0 \\
    0 & 0 & -1 & 0 & 1 \\
  \end{pmatrix},\qquad {}&
  \hess f_4 &= a^{-1}
  \begin{pmatrix}
      0 & 0 & 0 & 0 & 0 \\
      0 & 0 & 0 & 0 & 0 \\
      0 & 0 & 0 & 0 & 0 \\
      0 & 0 & 0 & 1 & 0 \\
      0 & 0 & 0 & 0 & 1 \\
  \end{pmatrix}.
\end{aligned}
\end{equation}

As we remarked in the main text and in Section~\ref{sm:sec:singular}, when using Eq.~\eqref{eq:mpd_singular} to find the marginal density $\mathcal{P}_{\hat{\theta}_{1}}(\mathcal{\theta}_{1})$, we will not make use of the parameterizations we derived earlier [Eqs.~\eqref{sm:eq:4bar_param_parallel} and \eqref{sm:eq:4bar_param_twisted}].
Instead, we need to first linearize the CV map at the singularity.
Since the CV we have chosen for the four-bar linkage is the angle $\theta_{1}$, the CV map that computes $\theta_{1}$ from the configuration vector $\bm{q}$ is $\hat{\theta}_{1}(\bm{q}) = \tan^{-1}(q_{5}/q_{4})$.\footnote{To be more rigorous, we should be using the two-argument variant of the inverse tangent, sometimes denoted as $\mathrm{atan2}(q_{5},q_{4})$ in numerical software, so that $\theta_{1}$ is in $(-\pi, \pi]$ instead of $(-\pi/2,\pi/2)$.  This is not an issue for the linearization since $\theta_{1}$ is small.}
This map has the Jacobian $\nabla\hat{\theta}_1 = \inmat{0 & 0 & 0 & 0 & a^{-1}}$ at the singularity.
Direct inspection reveals that $(\nabla\hat{\theta}_{1})\bm{v} = \bm{0}$ for all fast modes $\bm{v} \in (\ker\mathsf{C})^{\perp}$, justifying the usage of Eq.~\eqref{eq:mpd_singular} to find $\mathcal{P}_{\hat{\theta}_{1}}(\mathcal{\theta}_{1})$ when $\theta_{1} \to 0$.
This would not have been the case if, for instance, we had chosen the coordinate $q_{4}$ of joint~4 of the four-bar linkage (see Fig.~\ref{sm:fig:4bar}), as our CV.
In such a case $\nabla{q_{4}} = \inmat{0 & 0 & 0 & 1 & 0}$ and the fast modes, all of which are along the collinear bars with a nonzero fourth component, do not satisfy $(\nabla q_{4})\bm{v} = \bm{0}$.
Incidentally, in this case, the tangent vectors $\bm{t}$ at the singularity are such that $(\nabla q_{4})\bm{t} = \bm{0}$, which also makes $q_{4}$ a poor choice as the CV since it does not capture the slow modes along $\bm{t}$.
Continuing with $\theta_{1}$ as our CV, to evaluate the integral in Eq.~\eqref{eq:mpd_singular}, we choose the vector $\bm{u} \in \ker \mathsf{C}$ to be $\bm{u} = (0, 0, q_3, 0, q_{5})$, so that the vector $\bm{w}(\bm{u})$ is
\begin{equation}
  \begin{aligned}
    \bm{w}(\bm{u}) &= \left(\frac{1}{2}\bm{u}\trans \hess f_1 \bm{u},\, \frac{1}{2}\bm{u}\trans\hess f_2\bm{u},\, \frac{1}{2}\bm{u}\trans\hess f_3\bm{u},\, \frac{1}{2}\bm{u}\trans\hess f_4\bm{u} \right)\\
      &= \left[0,\,q_3^2/(2a),\,(q_3-q_5)^2/(2\lambda a),\,q_5^2/(2a)\right].
  \end{aligned}
  \label{sm:eq:4bar_w}
\end{equation}
There is only one self stress $\bm{\sigma}\in \ker\mathsf{C}\trans$ at the singularity, and it is $\bm{\sigma} = \left(-1/2, -1/2, 1/2, 1/2\right)$.
Using this in Eq.~\eqref{eq:mpd_singular} along with $(\nabla\hat{\theta}_{1})\bm{u} = a^{-1}q_{5}$ and the fact that $\det\,\mathsf{D}^\perp = 4\kappa^3$, we get%
\footnote{From the argument of the Dirac delta function, we see that the ``hyperplane'' $\Xi_{\xi}$ in Eq.~\eqref{eq:mpd_singular} is just the line along $q_{5} = a\theta_{1}$ in the $q_{3}$-$q_{5}$ plane.}
\begin{equation}
  \begin{aligned}
    \mathcal{P}_{\hat{\theta}_1}(\theta_1) 
                                           &\sim \frac{\lambda a}{2}\left(\frac{2\pi}{\beta\kappa}\right)^{3/2}\Biggl(\int_{-\infty}^{\infty} \dd{q}_3\, \dd{q}_{5}\, \delta(a^{-1}q_{5} - \theta_{1})\\
                                           &\phantom{\sim \frac{\lambda a}{2}\left(\frac{2\pi}{\beta\kappa}\right)^{3/2}}\quad\times\exp\left\{-\frac{\beta\kappa}{32\lambda^2a^2}\left[q_3 - q_5\right]^2\left[(\lambda-1)q_3 + (\lambda+1)q_5\right]^2\right\}\Biggr)\\
                                           &= \frac{\lambda a^2}{2}\left(\frac{2\pi}{\beta\kappa}\right)^{3/2}\int_{-\infty}^{\infty} \dd{q}_3\, \exp\left\{-\frac{\beta\kappa}{32\lambda^2a^2}\left[q_3 - a\theta_{1}\right]^2\left[(\lambda-1)q_3 + (\lambda+1)a\theta_{1}\right]^2\right\}.
  \end{aligned}
  \label{sm:eq:4bar_mpd_int}
\end{equation}
At this point, it is useful to revisit the convergence criteria for Eq.~\eqref{eq:mpd_singular}, i.e., the requirement that the term $\sum_{\bm{\sigma}} [\bm{\sigma}\cdot \bm{w}(\bm{u})]^{2}$ in the exponential of Eq.~\eqref{eq:mpd_singular} must only have isolated zeros. In the above equation, this term is $[q_3 - a\theta_{1}]^{2}[(\lambda-1)q_3 + (\lambda+1)a\theta_{1}]^{2}/(16\lambda^{2}a^{2})$, which has isolated zeros $q_{3} = a\theta_{1}$ and $q_{3} = (1+\lambda)a\theta_{1}/(1-\lambda)$, consistent with the convergence requirement.

To evaluate the integral in Eq.~\eqref{sm:eq:4bar_mpd_int}, we symmetrize the expression in the exponential by changing variables $q_3 \to q_3 - (\lambda-1)^{-1}a\theta_1$, which yields
\begin{equation}
  \begin{aligned}
    \mathcal{P}_{\hat{\theta}_1}(\theta_1) &\sim \frac{\lambda a^2}{2}\left(\frac{2\pi}{\beta\kappa}\right)^{3/2}\int_{-\infty}^{\infty} \dd q_3\, \exp\left\{-\frac{\beta\kappa(\lambda-1)^2}{32\lambda^2 a^2}\left[q_3^2 - \frac{\lambda^2 a^2\theta_1^2}{(\lambda-1)^2}\right]^2\right\}\\
                                           &= \lambda a^2\frac{(2\pi)^{3/2}}{(\beta\kappa)^{7/4}}\sqrt{\frac{\lambda{a}}{\abs{\lambda-1}}}\exp\left[-\frac{\beta\kappa\lambda^2a^2\theta_1^4}{32(\lambda-1)^2}\right]\int_{0}^{\infty} \dd x\, x^{-1/2}\exp\left(-\frac{1}{2}x^2 + \frac{\sqrt{\beta\kappa}\lambda a\theta_1^2}{4\abs{\lambda-1}}x\right)\\
                                           &= \lambda a^2\frac{(2\pi)^{3/2}}{(\beta\kappa)^{7/4}}\sqrt{\frac{\pi\lambda{a}}{\abs{\lambda-1}}}\exp\left[-\frac{\beta\kappa\lambda^2a^2\theta_1^4}{64(\lambda-1)^2}\right]D_{-1/2}\left(-\frac{\sqrt{\beta\kappa}\lambda a\theta_1^2}{4\abs{\lambda-1}}\right),
  \end{aligned}
  \label{sm:eq:4bar_0}
\end{equation}
where $D_{-1/2}(\cdot)$ is the parabolic cylinder function~\cite{olver2010}.

For the singularity at $\theta_{1} = \pm\pi$, we have $\bar{\bm{q}}^{*} = [\lambda a, (\lambda-1)a, 0, -a, 0]$ and we proceed with a similar calculation.
The dynamical matrix and the Hessians of the constraint functions at this point is identical to those at $\theta_{1} = 0$.
Hence, we choose the vector $\bm{u}$ as before, yielding the same $\bm{w}(\bm{u})$ as in Eq.~\eqref{sm:eq:4bar_w}.
However, the self stress at $\theta_1 = \pm\pi$ is different and it is $\bm{\sigma} = \left(-1/2, -1/2, -1/2, 1/2\right)$.
Using these results, we can evaluate the integral in Eq.~\eqref{eq:mpd_singular} as before to get
\begin{equation}
  \begin{aligned}
    \mathcal{P}_{\hat{\theta}_1}(\theta_1) &\sim \frac{\lambda a^2}{2}\left(\frac{2\pi}{\beta\kappa}\right)^{3/2}\int_{-\infty}^{\infty} \dd{q}_3\, \exp\left\{-\frac{\beta\kappa(\lambda+1)^2}{32\lambda^2 a^2}\left[q_3^2 - \frac{\lambda^2 a^2(\pi-\abs{\theta_1})^2}{(\lambda+1)^2}\right]^2\right\}\\
                                           &= \lambda a^2\frac{(2\pi)^{3/2}}{(\beta\kappa)^{7/4}}\sqrt{\frac{\pi\lambda{a}}{\lambda+1}}\exp\left[-\frac{\beta\kappa\lambda^2a^2(\pi-\abs{\theta_1})^4}{64(\lambda+1)^2}\right]D_{-1/2}\left[-\frac{\sqrt{\beta\kappa}\lambda a(\pi-\abs{\theta_1})^2}{4(\lambda+1)}\right].
  \end{aligned}
  \label{sm:eq:4bar_pi}
\end{equation}

Note that the marginal densities at the singular values of $\theta_{1}$, i.e., $\mathcal{P}_{\hat{\theta}_{1}}(0)$ and $\mathcal{P}_{\hat{\theta}_{1}}(\pm\pi)$, scale as $(\beta\kappa)^{-7/4}$.
Since $m = 4$ and the number of self stresses $s = 1$ at the singularities, this is consistent with the general scaling $\mathcal{P}_{\hat{\xi}}(\xi^{*}) \sim (\beta\kappa)^{-(m/2 - s/4)}$ for singular values $\xi^{*}$ of the CV [Eq.~\eqref{sm:eq:scaling}].

\subsection{Free energy}

Using the marginal densities we have found for various regimes of $\theta_{1}$ [Eqs.~\eqref{sm:eq:4bar_r}, \eqref{sm:eq:4bar_0}, and \eqref{sm:eq:4bar_pi}] we see that the free-energy difference of the four-bar linkage $\Delta\mathcal{A}_{\hat{\theta}_{1}}(\theta_1) = \mathcal{A}_{\hat{\theta}_{1}}(\theta_1) - \mathcal{A}_{\hat{\theta}_{1}}(0) = -\beta^{-1}\ln{\mathcal{P}_{\hat{\theta}_1}(\theta_1)} + \beta^{-1}\ln{\mathcal{P}_{\hat{\theta}_1}(0)}$ takes the form
\begin{equation}
  \Delta\mathcal{A}_{\hat{\theta}_{1}}(\theta_1) \sim
\begin{dcases}
  \beta^{-1}\left\{X^2\theta_1^4 - \ln\left[\dfrac{D_{-1/2}(-2X\theta_1^2)}{D_{-1/2}(0)}\right]\right\}, & \theta_1 \to 0\\
  \beta^{-1}\ln\left[X^{1/2}D_{-1/2}(0)\abs{\sin{\theta_1}}\right], & 0 \ll \abs{\theta_1} \ll \pi\\
  \beta^{-1}\left(X^2Y^2(\pi - \abs{\theta_1})^4 - \ln\left\{Y^{1/2}\dfrac{D_{-1/2}\left[-2XY(\pi - \abs{\theta_1})^2\right]}{D_{-1/2}(0)}\right\}\right), & \abs{\theta_1} \to \pi
\end{dcases}
  \label{sm:eq:4bar_free}
\end{equation}
where $X$ and $Y$ are positive dimensionless terms independent of $\theta_1$ and defined by
\begin{equation}
  X = \frac{\sqrt{\beta\kappa}\lambda a}{8\abs{\lambda-1}},\qquad
  Y = \left|\frac{\lambda-1}{\lambda+1}\right|.
\end{equation}
From Eq.~\eqref{sm:eq:4bar_free}, we also find the free-energy difference between the singular values $\theta_{1} = 0$ and $\theta_{1} = \pm\pi$ to be
\begin{equation}
  \mathcal{A}_{\hat{\theta}_{1}}(\pm\pi) - \mathcal{A}_{\hat{\theta}_{1}}(0) \sim -\beta^{-1}\ln{Y^{1/2}} = \frac{1}{2}\beta^{-1}\ln\left|\frac{\lambda + 1}{\lambda - 1}\right|,
\end{equation}
which is a purely geometric quantity.
This is exactly what we expect based on how the marginal density scales at a singular value [Eq.~\eqref{sm:eq:scaling}], which shows that the free-energy difference must be a purely geometric quantity if the singular states support the same number of self stresses $s$.

\altsection{Double-well structures}

\begin{figure}
  \begin{center}
    \includegraphics{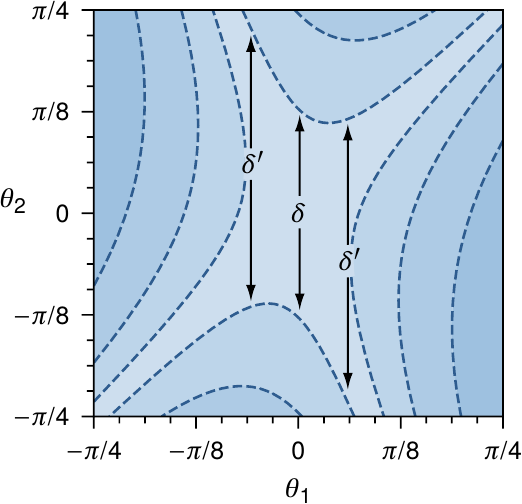}
  \end{center}
  \caption{The level sets of the energy in Eq.~\eqref{sm:eq:energy34} around $(\theta_{1}, \theta_{2}) = (0, 0)$ for $a = \kappa = 1$ and $\lambda = 2$.  The width of the energy sublevel sets is larger along small nonzero values of $\theta_{1}$ compared to $\theta_{1} = 0$ (e.g., $\delta' > \delta$).}
  \label{sm:fig:energy34}
\end{figure}

To intuitively see why the free-energy landscape of the four-bar linkage has a double-well structure centered around the singular values, consider the projection of its total energy in the space of the angles $\theta_{1}$ and $\theta_{2}$.
The projected energy can be found by setting the lengths of all bars equal to their natural lengths except for bar~3--4, which we assume to have an energy $\phi(\ell) = \kappa(\ell^{2} - \bar{\ell}^{2})^{2}/(8\bar{\ell}^{2})$.
(The exact form of the energy is irrelevant as long as it has a minimum at $\ell = \bar{\ell}$.)
Writing $\ell$ in terms of the angles $\theta_{1}, \theta_{2}$ we get
\begin{equation}
  \phi(\theta_{1}, \theta_{2}) = \frac{\kappa a^{2}}{8\lambda^{2}}\left[(\lambda + \cos\theta_{2} - \cos\theta_{1})^{2} + (\sin\theta_{2} - \sin\theta_{1})^{2} - \lambda^{2}\right]^{2}\,.
  \label{sm:eq:energy34}
\end{equation}
Figure~\ref{sm:fig:energy34} shows the level sets of $\phi(\theta_{1}, \theta_{2})$ near $(\theta_{1}, \theta_{2}) = (0, 0)$.
For $\abs{\theta_{1}} > 0$, there are two ground states in the CV level set $\hat{\theta}_{1}^{-1}(\theta_{1})$, which is a straight line parallel to the $\theta_{2}$ axis in the $\theta_{1}$--$\theta_{2}$ space.
Because of the extra softness of the linkage near the singularity, this means that for small nonzero values of $\theta_{1}$, there are more thermodynamically favorable states in $\hat{\theta}_{1}^{-1}(\theta_{1})$ compared to $\hat{\theta}_{1}^{-1}(0)$, which is the CV level set at the singular value $\theta_{1}=0$.
This is evidenced by the fact that for any given value $E > 0$, there are more points in the energy sublevel set $\left\{(\theta_{1}, \theta_{2}): \phi(\theta_{1}, \theta_{2}) \leq E\right\}$ along small nonzero values of $\theta_{1}$ than $\theta_{1} = 0$ (see Fig.~\ref{sm:fig:energy34}).
The net increase in the number of thermodynamically favorable states near small nonzero values of $\theta_{1}$ lowers the free energy at those values, giving the landscape a double-well appearance.

We also remark that since an asymptotic expression for the free energy is available for the four-bar linkage, we can explicitly show the double-well nature of $\Delta\mathcal{A}_{\hat{\theta}_{1}}(\theta_{1})$ and find locations of its minima.
For example, expanding Eq.~\eqref{eq:4bar_0} in $\theta_{1}$ around $\theta_{1} = 0$, we see that
\begin{equation}
  \Delta\mathcal{A}_{\hat{\theta}_{1}}(\theta_{1}) \sim \left[1 + 8\pi^{2}\Gamma^{-4}\left(\tfrac{1}{4}\right)\right]X^{2}\theta_{1}^{4} - 4\pi\Gamma^{-2}\left(\tfrac{1}{4}\right)X\theta_{1}^{2},\label{sm:eq:4barwell}
\end{equation}
which is the equation for a double well, with $\Gamma(\cdot)$ being the gamma function.
Expanding $\Delta\mathcal{A}_{\hat{\theta}_{1}}(\theta_{1})$ to $\mathcal{O}(\abs{\theta_{1}}^{6})$ we see that the two double-well minima are approximately at\footnote{We expand $\Delta\mathcal{A}_{\hat{\theta}_{1}}(\theta_{1})$ to $\mathcal{O}(\abs{\theta_{1}}^{6})$ instead of $\mathcal{O}(\abs{\theta_{1}}^{4})$ since there are higher-order corrections to Eq.~\eqref{sm:eq:4barwell} that make an $\mathcal{O}(\abs{\theta_{1}}^{4})$ estimation less accurate.}
\begin{equation}
  \theta_{1}^{\text{min}} \approx \pm \sqrt{\frac{\Gamma^{2}\left(\frac{1}{4}\right)\left\{8\pi^{2} + \Gamma^{4}\left(\frac{1}{4}\right) - \sqrt{\left[8\pi^{2} + \Gamma^{4}\left(\frac{1}{4}\right)\right]^{2} - (16\pi^{2})^{2}}\right\}}{64\pi^{3}X}}.
  \label{sm:eq:thetamin}
\end{equation}
Since $X \sim \sqrt{\beta}$, this also shows that as $\beta$ increases, $\theta_{1}^{\text{min}}$ shifts closer to $0$.

\section{Triangulated Origami}
\label{sm:sec:origami}

\subsection{Body frame}

\begin{figure}
  \begin{center}
    \includegraphics{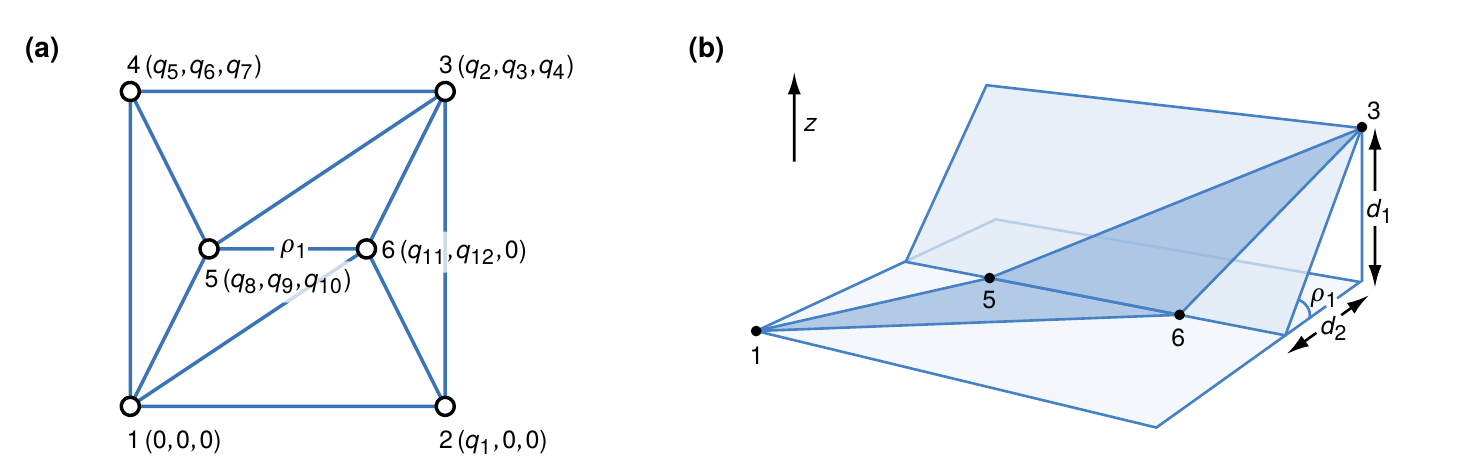}
  \end{center}
  \caption{(a) A triangulated origami as a bar-joint mechanism illustrating the coordinates of the joints in the body frame. When the origami is flat, the external vertices lie on the corners of a unit square and the internal vertices 5 and 6 have coordinates $(1/4,1/2,0)$ and $(3/4,1/2,0)$. (b) Finding the CV (i.e., fold angle $\rho_{1}$) using simple geometry.}
  \label{sm:fig:origami}
\end{figure}

To remove the rigid motions, we transform to a body frame attached to the origami with joint~1 at the origin, bar~1--2 lying along the $x$ axis, and bar~2--6 constrained to move on the $xy$ plane [see Fig.~\ref{sm:fig:origami}(a)].
The origami is made out of $N = 6$ joints and $m = 11$ bars and its configuration vector $\bm{q} \in \mathbb{R}^{12}$ in the body frame is $\bm{q} = (q_{1}, q_{2}, \ldots, q_{12})$.
All orientations of the origami in the lab frame can be fully described by the two spherical polar angles that uniquely give the orientation of bar~1--2 and an azimuthal angle that gives the overall rotation of the origami about bar~1--2~\cite{herschbach1959}.
After integrating over the coordinates of joint~1 (i.e., the translational coordinates) and the three angles, and dropping constant factors, the overall Jacobian factor involved in the transformation from the lab to the body frame is $I(\bm{q}) = \abs{q_1^2 q_{12}}$.

\subsection{Branch parameterization}

Since the shape space of the origami (and other larger mechanisms) is not amenable to analytical parameterization, we have to parameterize it numerically.
We first express the tangent to the shape space at the flat state, $\bm{t}_{0} \in \ker \mathsf{C}$, in terms of its unknown components in the basis of the $(N - 3)$ out-of-plane displacement vectors of the joints.\footnote{Note that three of the $N$ joints are always constrained to move on a coordinate plane of the local Cartesian body frame.  This means that only $(N - 3)$ joints have out-of-plane displacements, and it is these displacement vectors that span $\ker\mathsf{C}$.}
An origami made by triangulating a square and having $N$ joints is expected to have at least $(N - 4)$ states of self stress $\bm{\sigma} \in \ker\mathsf{C}\trans$ when it is flat~\cite{chen2018}.
To find the components of $\bm{t}_{0}$, we then solve the $(N - 4)$ coupled quadratic equations~\cite{tarnai2001,chen2018} [see Eq.~\eqref{sm:eq:tangentcone}]
\begin{equation}
  \bm{\sigma}\cdot\bm{w}(\bm{t}_{0}) = \bm{0},\quad \text{for } \bm{\sigma} \in \ker{\mathsf{C}\trans}
  \label{sm:eq:quadratic}
\end{equation}
along with the normalization condition $\Abs{\bm{t}_{0}} = 1$.
Here the $i$th component of the vector $\bm{w}(\bm{t}_{0}) \in \mathbb{R}^{m}$ is $\frac{1}{2}\bm{t}_{0}\trans \hess f_{i} \bm{t}_{0}$, as in the main text.

Numerical evidence~\cite{chen2018} suggests that one would get $2^{N-4}$ unique tangent vectors---each corresponding to a particular branch---by solving the above quadratic equations.
(Also see the related discussion regarding the solution space of these equations in Section~\ref{sm:sec:convergence}.)
Once a tangent vector to a branch at the flat state is selected, the rest of the branch can be parameterized by numerically solving the differential equation $\dd\bm{q}/\dd{h} = \bm{t}$, where $h$ is the arc length along a branch and $\bm{t}$ is the unit tangent vector to the branch at $\bm{q}$.
If the $k$th point on the branch is $\bm{q}_k \in \mathbb{R}^{n}$, the next point $\bm{q}_{k+1}$ is then found by solving $f(\bm{q}_{k+1}) = \bm{0}$, e.g., using the Gauss--Newton method.
As the initial guess we take $\bm{q}_{k+1} = \bm{q}_{k} + \Delta h\,\bm{t}_{k}$, where $\Delta h$ is the step length along a branch.
The tangent vector $\bm{t}_k$ for $k \ne 0$ can be found by numerically computing $\ker\mathsf{C}$ at $\bm{q}_k$.
Since the tangent vectors obtained this way does not preserve direction in general, we should also multiply each $\bm{t}_k$ that is found with the sign of its dot product with the previous tangent vector $\bm{t}_{k-1}$.

On successive repetition of the above steps starting at the flat state $(\bm{q}_0, \bm{t}_0)$, we get $\bm{q}$ as a function of the arc length $h$.
To reparameterize the branch using the CV, which is the fold angle $\rho_{1}$ of fold 5--6, we first (linearly) interpolate between the arc-length parameterized points to obtain a set of points that are uniformly spaced in $\rho_{1}$.
We then refine the interpolated points by solving $f(\bm{q}) = \bm{0}$ with the interpolated points as the initial guess.
The interpolation/refinement steps can be repeated as many times as required to achieve the desired accuracy goal.
Once the parameterization is complete, the induced metric along a branch can be computed by approximating the derivatives using difference quotients.
Note that this parameterization is only used in conjunction with Eq.~\eqref{eq:mpd_regular}.

\subsection{Marginal probability densities}

\subsubsection{Regular values}

Similar to the calculation for the four-bar linkage, we first write down the constraint map $f: \mathbb{R}^{12} \to \mathbb{R}^{11}$ in the body frame and find the compatibility matrix $\mathsf{C} = \nabla f$.
Once all four branches of the shape space have been numerically parameterized in terms of the CV $\rho_{1}$, the marginal probability density $\mathcal{P}_{\hat{\rho}_{1}}(\rho_{1})$ can be computed using Eq.~\eqref{eq:mpd_regular}.
Here we remark that instead of computing $\det\,\mathsf{D}^{\perp}$ by finding the nonzero eigenvalues of $\mathsf{D}$, it is more convenient to calculate it using the fact that $\det\,\mathsf{D}^{\perp} = \det\,\mathsf{K}\mathsf{C}\mathsf{C}\trans$ at regular points.
This gives $\mathcal{P}_{\hat{\rho}_{1}}(\rho_{1})$ for all $\rho_{1}$ far from the singular value (i.e., for $\abs{\rho_{1}} \gg 0$).

\subsubsection{Singular value}

Our goal here is to use Eq.~\eqref{eq:mpd_singular} to find the marginal density $\mathcal{P}_{\hat{\rho}_{1}}(\rho_{1})$ as $\rho_{1} \to 0$.
Since the calculation is similar in spirit to the case of the four-bar linkage, we only present the key steps here.
As before, we numerically compute $\det\,\mathsf{D}^{\perp}$ from the nonzero eigenvalues of $\mathsf{D}$ at the singularity.
The next step is to linearize the CV map so that the integral in Eq.~\eqref{eq:mpd_singular} can be evaluated.

One can always numerically compute the fold angle $\rho_{1}$ as the angle between the normals to the faces that share fold 5--6.
However, for linearizing the CV map, we need to find an expression that is more tractable analytically.
A moment's thought [and Fig.~\ref{sm:fig:origami}(b)] shows that the CV map can be written\footnote{To assign a sign to the fold angle, we first choose a unique normal to face 1--5--6 such that it coincides with the positive $z$ axis when the origami is flat.  Signs are then chosen so that a mountain fold (as perceived by looking downwards along this normal) has a positive fold angle, e.g., the angle in Fig.~\ref{sm:fig:origami}(b) is negative since the fold is a valley fold.} as $\hat{\rho}_{1}(\bm{q}) = -\tan^{-1}[d_{1}(\bm{q})/d_{2}(\bm{q})]$.
Here $d_{1}(\bm{q})$ is the perpendicular distance from joint~3 to the plane containing face 1--5--6, and $d_{2}(\bm{q})$ is the perpendicular distance from the projection of joint~3 on this plane to the line along fold 5--6.
A straightforward calculation gives
\begin{equation}
  d_{1}(\bm{q}) = \frac{q_{10}(q_{2}q_{12} - q_{3}q_{11}) + q_4(q_{9}q_{11} - q_{8}q_{12})}{\left|q_{10}^2 q_{11}^2+q_{10}^2 q_{12}^2+(q_9 q_{11}-q_8 q_{12})^2\right|^{1/2}}.
\end{equation}
At the singular flat state $\bar{\bm{q}}^{*}$, we have $d_{1}(\bar{\bm{q}}^{*}) = 0$ and $d_{2}(\bar{\bm{q}}^{*}) = 0.5$, using which we find the Jacobian of the CV map to be
\setcounter{MaxMatrixCols}{20}
\begin{equation}
\begin{aligned}
  \nabla\hat{\rho}_{1}(\bar{\bm{q}}^{*}) = -\frac{d_{2}(\bar{\bm{q}}^{*})\nabla d_{1}(\bar{\bm{q}}^{*}) - d_{1}(\bar{\bm{q}}^{*})\nabla d_{2}(\bar{\bm{q}}^{*})}{d^{2}_{1}(\bar{\bm{q}}^{*}) + d^{2}_{2}(\bar{\bm{q}}^{*})} &= -\frac{\nabla d_{1}(\bar{\bm{q}}^{*})}{d_{2}(\bar{\bm{q}}^{*})}\\
  &= \inmat{0 & 0 & 0 & -2 & 0 & 0 & 0 & 0 & 0 & 2 & 0 & 0}.
\end{aligned}
\end{equation}
Here we see that one can linearize the CV map without needing an explicit expression for $d_{2}(\bm{q})$.
Also, it is easy to verify that $(\nabla\hat{\rho}_{1})\bm{v} = \bm{0}$ for all fast modes $\bm{v} \in (\ker\mathsf{C})^{\perp}$, enabling us to use Eq.~\eqref{eq:mpd_singular} to find the marginal density as $\rho_{1} \to 0$.

As the basis for $\ker\mathsf{C}$ we choose the out-of-plane displacements of the joints in the body frame and write the vector $\bm{u} \in \ker \mathsf{C}$ in terms of the $z$ coordinates $q_{4}$, $q_{7}$, and $q_{10}$ of the joints.
This yields $\nabla\hat{\rho}_{1}\cdot\bm{u} = 2(q_{10} - q_{4})$.
It also follows that the hyperplane $\Xi_{\xi}$ is the plane defined by $2(q_{10} - q_{4}) = \rho_{1}$ in the $q_{4}$-$q_{7}$-$q_{10}$ space and we can parameterize the points on $\Xi_{\xi}$ using either $(q_{4}, q_{7})$ or $(q_{7}, q_{10})$.
Once we write down the vector $\bm{w}(\bm{u})$ and find the self stresses $\bm{\sigma}$, we have all the ingredients to use Eq.~\eqref{eq:mpd_singular}.
However, the integral that one is left with is not amenable to analytical integration, and so we have to resort to numerical quadrature.
We do this by using a simple Monte Carlo integration scheme with $10^{9}$ sample points for each value of $\rho_{1}$ so that the maximum error is below $10^{-3}$.
One also gets near-identical results with Mathematica's numerical integrator using a global adaptive method.\footnote{\url{https://reference.wolfram.com/language/ref/NIntegrate.html}}

\subsection{Free energy}

The free-energy difference $\Delta\mathcal{A}_{\hat{\rho}_{1}}(\rho_{1}) = \mathcal{A}_{\hat{\rho}_{1}}(\rho_{1}) - \mathcal{A}_{\hat{\rho}_{1}}(0)$ for all values of $\rho_{1}$ can be easily computed from the corresponding probability densities using Eq.~\eqref{eq:free_energy}.
Similar to the four-bar linkage, the free energy of the origami also has a double-well appearance around the singular value $\rho_{1} = 0$.
This is again because of the branched nature of its shape space and the increased softness near the singularity.
\begin{figure}
  \begin{center}
    \includegraphics[scale=1.0]{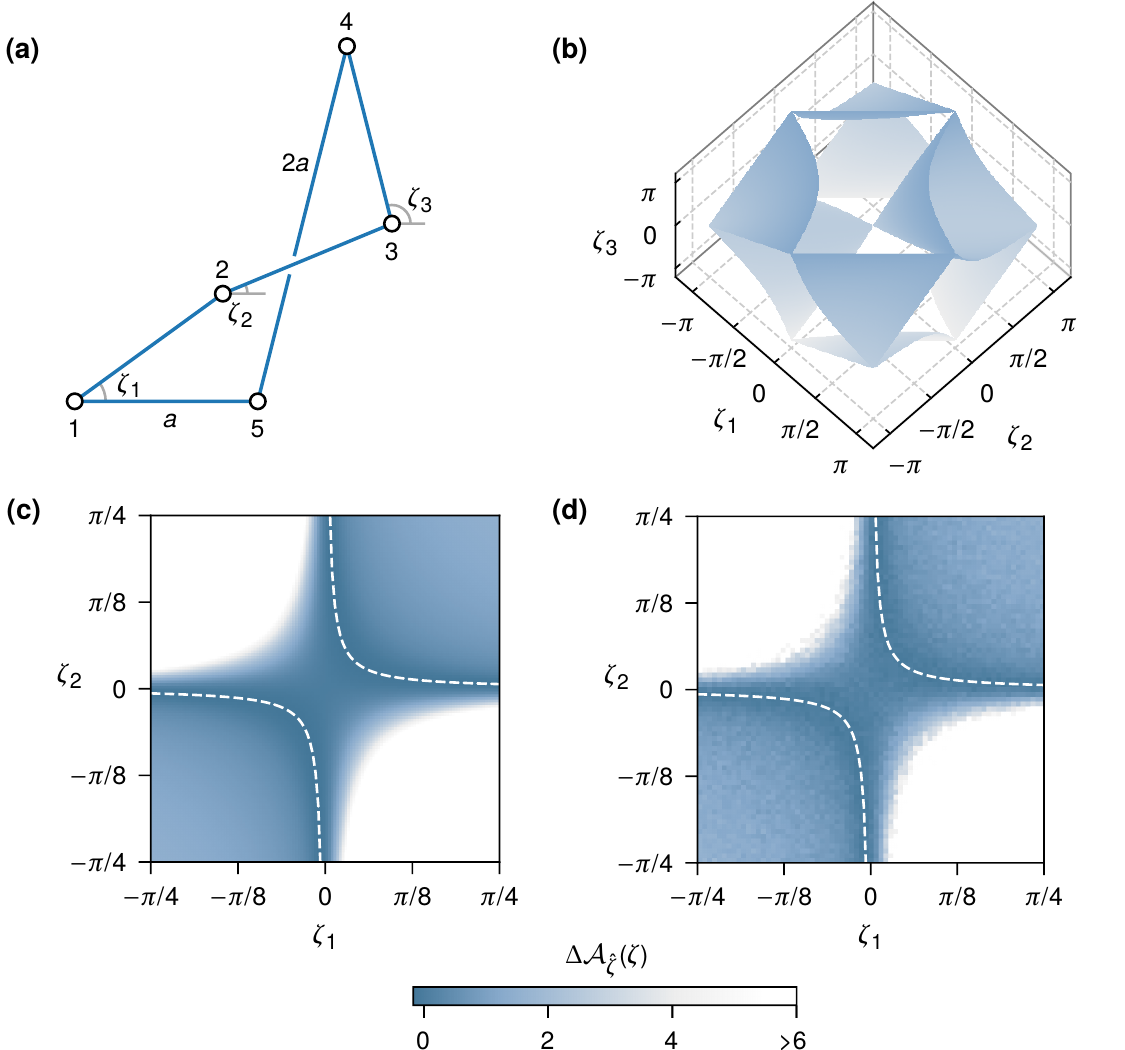}
  \end{center}
  \caption{(a) The five-bar linkage and (b) its shape space visualized in terms of the angles $\zeta_{1}, \zeta_{2},$ and $\zeta_{3}$.  These angles are measured counterclockwise from an axis parallel to bar~1--5.  Free-energy profile (in units of $\beta^{-1})$ at $\beta = 10^{4}$ around the singular value $\zeta^{*} = (0, 0)$ from (c) theory [Eq.~\eqref{sm:eq:5bar}] and (d) simulations, with the dashed curves depicting the approximate bottom of the hyperbolic valley [Eq.~\eqref{sm:eq:5barvalley}].}
  \label{sm:fig:5bar}
\end{figure}

\section{Planar Five-Bar Linkage}
\label{sm:sec:5bar}

The example mechanisms we have considered so far have one-dimensional shape spaces.
In this section, we illustrate the application of our formalism to a mechanism with a two-dimensional shape space, namely the planar five-bar linkage~\cite{mermoud2000,curtis2007} illustrated in Fig.~\ref{sm:fig:5bar}(a).
The five-bar linkage we consider is made out of four bars of equal length $a$ and a fifth bar of length $2a$.
The shape space of the five-bar linkage [Fig.~\ref{sm:fig:5bar}(b)] when visualized in the space of the angles $\zeta_{1}, \zeta_{2}$, and $\zeta_{3}$ appears as a two-dimensional surface with four isolated singularities,\footnote{At $(\zeta_{1}, \zeta_{2}, \zeta_{3}) = (0, 0, 0), (0, \pm\pi, \pm\pi), (\pm\pi, 0, \pm\pi), \text{ and } (\pm\pi, \pm\pi, 0)$.} all of which correspond to configurations where the bars become collinear and support a state of self stress.
Since force balance in self-stressed states of a planar polygonal linkage requires its bars to be collinear~\cite{farber2008}, it is not surprising that these singularities are isolated, and in their neighborhoods, the shape space is very nearly a double cone~\cite{kapovich1995,mermoud2000}.
Similar isolated singularities are also seen in the shape spaces of origami that have self-stressed flat states~\cite{berry2020}.

For the sake of brevity and to avoid cluttering this SM with qualitatively similar results, we only discuss the nature of the free-energy landscape around the singularity at $(\zeta_{1}, \zeta_{2}, \zeta_{3}) = (0, 0, 0)$.
Since the shape space of the five-bar linkage is two dimensional, we choose a similarly two-dimensional CV, $\zeta = (\zeta_{1}, \zeta_{2})$.
Next, we employ Eq.~\eqref{eq:mpd_singular} to find the asymptotic free-energy difference $\Delta \mathcal{A}_{\hat{\zeta}}(\zeta)$, choosing the singular value $\zeta^{*} = (0,0)$ as the point of zero free energy.
Note that for doing this, we do not require an explicit parameterization of the linkage's shape space in terms of $\zeta$.  Such a parameterization is only required if we want to use Eq.~\eqref{eq:mpd_regular} to find the free energy far from $\zeta^{*} = (0, 0)$ using the harmonic approximation.
Since the other details are very similar to the previous calculations for the four-bar linkage and the triangulated origami, we just quote the final result:
\begin{equation}
  \Delta\mathcal{A}_{\hat{\zeta}}(\zeta) \sim \beta^{-1}\left\{Z^{2}\zeta_{1}^{2}\zeta_{2}^{2} - \ln\left[\frac{D_{-1/2}(-2Z\zeta_{1}\zeta_{2})}{D_{-1/2}(0)}\right]\right\},\label{sm:eq:5bar}
\end{equation}
where $Z$ is a positive dimensionless term defined by $Z = \sqrt{\beta\kappa}a/(2\sqrt{5})$ [cf.~Eq.~\eqref{eq:4bar_0}].
A comparison between the theoretical and numerical results [Figs.~\ref{sm:fig:5bar}(c) and \ref{sm:fig:5bar}(d)] shows very good agreement between the two.
Similar to the four-bar linkage and the triangulated origami, the effective free energy minimum of the five-bar linkage is not at the singular value $\zeta^{*} = (0, 0)$.
Expanding $\Delta\mathcal{A}_{\hat{\zeta}}(\zeta)$ to $\mathcal{O}(\zeta_{1}^{3}\zeta_{2}^{3})$ around $\zeta^{*}$ we see that the bottom of the free-energy valley near the singular value [white dashed curves in Figs.~\ref{sm:fig:5bar}(c) and \ref{sm:fig:5bar}(d)] is approximately defined by the equation
\begin{equation}
  \zeta_{1}\zeta_{2} = \frac{\Gamma^{2}\left(\frac{1}{4}\right)\left\{8\pi^{2} + \Gamma^{4}\left(\frac{1}{4}\right) - \sqrt{\left[8\pi^{2} + \Gamma^{4}\left(\frac{1}{4}\right)\right]^{2} - (16\pi^{2})^{2}}\right\}}{64\pi^{3}Z},
  \label{sm:eq:5barvalley}
\end{equation}
which is that of a rectangular hyperbola [cf.~Eq.~\eqref{sm:eq:thetamin}].

\section{Permanently singular mechanisms}
\label{sm:sec:permanent}

\begin{figure}
  \begin{center}
    \includegraphics[scale=1.0]{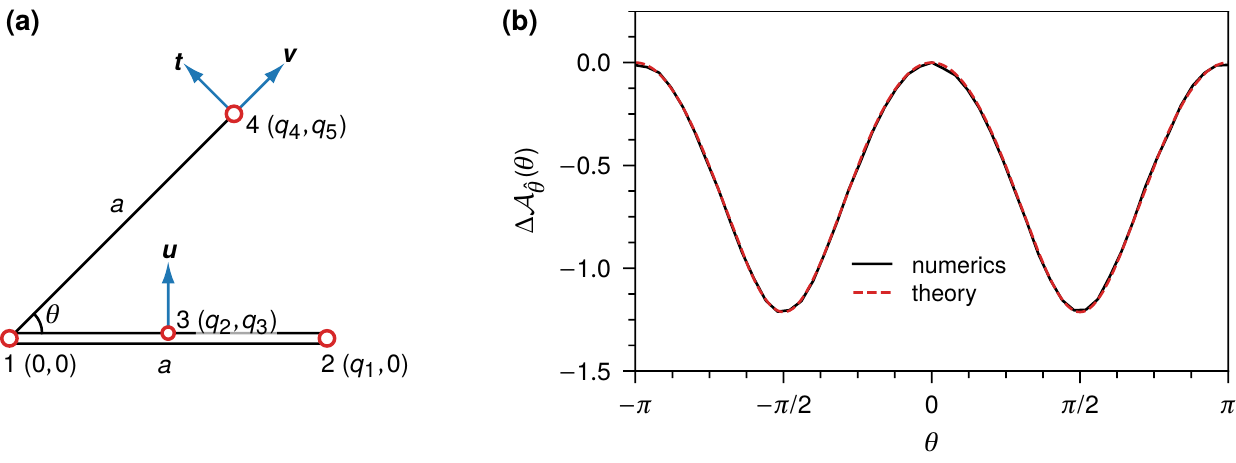}
  \end{center}
  \caption{(a) A planar mechanism with four joints and four bars in a body frame attached to the mechanism.  A bar of length $a$ connects joints~1 and~2, and two other bars of length $\frac{1}{2}a$ connect joint~3 with joints~1 and~2, which results in a state of self stress between joints 1, 2, and 3~\cite{calladine1978} for all values of $\theta$. Here $\bm{t}$ is the zero mode corresponding to tangential motion on the shape space $\Sigma$, $\bm{u}$ is a singular zero mode associated with the self stress, and $\bm{v}$ is one of the three vibrational modes.  (b) Free-energy difference $\Delta\mathcal{A}_{\hat{\theta}}(\theta)$ [Eq.~\eqref{sm:eq:trimer_free}] in units of $\beta^{-1}$ for a hypothetical $\theta$-dependent stiffness $\kappa(\theta) = \kappa_{0}(1 + \cos^{2}{\theta})$, chosen to verifying scaling. For constant stiffness, $\Delta\mathcal{A}_{\hat{\theta}}(\theta)$ is zero and the free-energy landscape is flat (unlike in the case of isolated singularities).}
  \label{sm:fig:trimer}
\end{figure}

Our discussion so far has been concerned with mechanisms with isolated singularities.
In such mechanisms, the constraint map $f$ drops rank only at the singularities and has full rank everywhere else.
Now consider the mechanism shown in Fig.~\ref{sm:fig:trimer}(a).
The bars connecting joints~1, 2, and 3 of this mechanism are in a permanent state of self stress, irrespective of the value of the angle $\theta$.
By direct inspection we see that the one-dimensional shape space $\Sigma$ of this mechanism can be parameterized in terms of the internal angle $\theta$ using $\psi(\theta) = (a, \frac{1}{2}a, 0, a\cos{\theta}, a\sin{\theta})$.
This equation defines a smooth circle in $\mathbb{R}^{5}$, which does not have any ``visible'' singularities.
However, on inserting this parameterization into the compatibility matrix $\mathsf{C} = \nabla f$ and the dynamical matrix $\mathsf{D} = \mathsf{C}\trans\mathsf{K}\mathsf{C}$ we find (assuming that all bars have stiffness $\kappa$)
\begin{equation}
  \mathsf{C} =
  \begin{pmatrix}
    1 & 0 & 0 & 0 & 0 \\
    1 & -1 & 0 & 0 & 0 \\
    0 & 1 & 0 & 0 & 0 \\
    0 & 0 & 0 & \cos \theta & \sin \theta
  \end{pmatrix},
  \quad
  \mathsf{D} =
  \kappa
  \begin{pmatrix}
    2 & -1 & 0 & 0 & 0 \\
    -1 & 2 & 0 & 0 & 0 \\
    0 & 0 & 0 & 0 & 0 \\
    0 & 0 & 0 & \cos ^2\theta & \sin \theta  \cos \theta \\
    0 & 0 & 0 & \sin \theta \cos \theta & \sin ^2\theta
  \end{pmatrix}.
\end{equation}
Clearly, $\mathsf{C}$ is rank deficient irrespective of the value of $\theta$ and $\mathsf{D}$ has two zero modes: $\bm{t} = \dd\psi/\dd\theta \in T_{\bar{\bm{q}}}\Sigma$ corresponding to tangential motion on $\Sigma$ and a singular zero mode $\bm{u} = (0, 0, 1, 0, 0)$ associated with the self stress [see Fig.~\ref{sm:fig:trimer}(a)].
Hence, even though the shape space here is a smooth manifold, the presence of the singular zero mode causes the harmonic approximation to break down everywhere on $\Sigma$.
This should be contrasted with the case of isolated singularities, where such singular modes appear only at the singularities of $\Sigma$.
The results that we have derived so far (namely, Eqs.~\ref{eq:mpd_regular} and \ref{eq:mpd_singular}) will not let us analyze permanently singular mechanisms~\cite{muller2019,wu2020} such as the one in Fig.~\ref{sm:fig:trimer}(a).
However, such mechanisms have been considered in the context of colloidal clusters~\cite{kallus2017}.
It is thus instructive to rederive these results and compare them with our results for mechanisms with isolated singularities.

Consider again a mechanism whose shape space $\Sigma$ is defined as the zero level set of a constraint map $f: \mathbb{R}^{n} \to \mathbb{R}^{m}$ with the compatibility matrix $\mathsf{C} = \nabla f$.
When there are $s$ permanent states of self stress, out of the $n - m + s$ zero modes that belong to $\ker\mathsf{C}$, there are $n-m$ zero modes that belong to the tangent space $T_{\bar{\bm{q}}}\Sigma$, and the remaining $s$ zero modes are the singular zero modes.
We can thus write $\ker\mathsf{C}(\bar{\bm{q}}) = T_{\bar{\bm{q}}}\Sigma \oplus \mathcal{S}$ for all $\bar{\bm{q}} \in \Sigma$.
Here $\mathcal{S}$ is the subspace of the singular zero modes at $\bar{\bm{q}}$, defined as the orthogonal complement of $T_{\bar{\bm{q}}}\Sigma$ in $\ker\mathsf{C}$.\footnote{Since the zero modes are degenerate, the singular zero modes obtained by computing $\ker\mathsf{C}$ (or $\ker\mathsf{D}$) need not be orthogonal to $T_{\bar{\bm{q}}}\Sigma$.  Hence, in writing $\ker\mathsf{C} = T_{\bar{\bm{q}}}\Sigma \oplus \mathcal{S}$, we are \emph{defining} a singular mode to be one that belongs to $\ker\mathsf{C}$, but is orthogonal to $T_{\bar{\bm{q}}}\Sigma$. Clearly, such zero modes cannot be extended to a smooth deformation of the mechanism.}
Note that such a decomposition of $\ker\mathsf{C}$ into two vector subspaces is not possible when $\Sigma$ has isolated singularities (where multiple branches cross) for two reasons: (i) singular zero modes exist only at the singularities of $\Sigma$ and (ii) at these singularities, there is no well-defined tangent space $T_{\bar{\bm{q}}}\Sigma$.

Since the subspace $\mathcal{S}$ of singular zero modes can be identified for all points in $\Sigma$, we can expand the energy to quartic order along $\bm{u} \in \mathcal{S}$ using Eq.~\eqref{eq:energy_singular}.
(This would not have been possible for isolated singularities since Eq.~\eqref{eq:energy_singular} is valid only when the expansion is around a singularity.)
To derive the asymptotic marginal density $\mathcal{P}_{\hat{\theta}}(\theta)$, we can then proceed similar to the derivation of Eq.~\eqref{eq:mpd_regular}.
As before, we pick coordinates associated with the mechanism's internal degrees of freedom as our CV $\xi$ and assume that $\Sigma$ can be parameterized using $\xi$.
We expand both the terms in the exponent of Eq.~\eqref{sm:eq:prob_delta} around each ground state $\bar{\bm{q}} \in \hat{\xi}^{-1}(\xi)$ after setting $\bm{q} \to \bar{\bm{q}} + \bm{t} + \bm{u} + \bm{v}$, where $\bm{t} \in T_{\bar{\bm{q}}}\Sigma$ is a zero mode that can be extended to a smooth deformation of the mechanism, $\bm{u} \in \mathcal{S}$ is a singular zero mode, and $\bm{v} \in (\ker\mathsf{C})^{\perp}$ is a fast mode.
We also choose the columns of $\nabla\psi(\xi)$ as the basis for $T_{\bar{\bm{q}}}\Sigma$, with $\psi: \mathbb{R}^{n-m} \to \mathbb{R}^{n}$ being the parameterization of $\Sigma$ near $\bar{\bm{q}}$.
Integrating over the components of $\bm{t}$ yields
\begin{equation}
  \begin{aligned}
    \mathcal{P}_{\hat{\xi}}(\xi) &\sim I(\xi)\left(\frac{2\pi}{\beta}\right)^{(m - s)/2} \left|\det\,[\nabla\psi(\xi)]\trans\nabla\psi(\xi)\right|^{1/2} \int_{\mathcal{S}} \dd\bm{u} \int_{(\ker\mathsf{C})^{\perp}} \dd\bm{v}\, \exp\left[-\frac{1}{2}\beta\kappa\Abs{\mathsf{C}\bm{v} + \bm{w}(\bm{u})}^{2}\right]\\
                                 &= I(\xi)\left(\frac{2\pi}{\beta}\right)^{(m - s)/2}(\beta\kappa)^{-s/4} \left|\frac{\det\,[\nabla\psi(\xi)]\trans\nabla\psi(\xi)}{\det\,\mathsf{D}^{\perp}(\xi)}\right|^{1/2} \int_{\mathbb{R}^{s}} \dd\bm{x}\, \exp\left\{-\frac{1}{2}\sum_{\bm{\sigma}}[\bm{\sigma}\cdot\bm{w}(\bm{x})]^{2}\right\}.
  \end{aligned}
  \label{sm:eq:mpd_perm}
\end{equation}
In the last step, we have integrated over the fast modes in $(\ker\mathsf{C})^{\perp}$ and used the results in Eqs.~\eqref{sm:eq:mpd_gaussian}--\eqref{sm:eq:mpd_proj}, with $\det\mathsf{D}^{\perp}$ being the product of the $m-s$ nonzero eigenvalues of the dynamical matrix $\mathsf{D}$ at $\psi(\xi)$.
Also, after picking an orthonormal basis for $\mathcal{S}$ and writing $\bm{u} = \mathsf{A}\bm{x}$, we can rescale the components $\bm{x} \to (\beta\kappa)^{-1/4}\bm{x}$ to extract all $\beta$ and $\kappa$ dependence as the term $\sum_{\bm{\sigma}} [\bm{\sigma}\cdot \bm{w}(\bm{x})]^{2}$ is a homogeneous quartic polynomial in the components of $\bm{x}$.
Equation~\eqref{sm:eq:mpd_perm} is a rederivation of the integrand in Eq.~(14) of Ref.~\cite{kallus2017} and although it looks very similar to Eq.~\eqref{eq:mpd_singular} of the main text, it is fundamentally different.
The similarities arise due to the fact that in deriving both Eqs.~\eqref{sm:eq:mpd_perm} and \eqref{eq:mpd_singular}, we used the same quartic-order expansion of the energy.
Nonetheless, it is worthwhile to compare the two equations.
We can identify four major differences.

\paragraph{Integration domain.}
The integration domain in Eq.~\eqref{sm:eq:mpd_perm} is $\mathcal{S}$, the subspace of singular zero modes, which exists for all configurations of a permanently singular mechanism.
In comparison, when the mechanism has isolated singularities, singular zero modes arise only at these singularities.
Furthermore, the domain of integration in Eq.~\eqref{eq:mpd_singular} is a CV-dependent hyperplane $\Xi_{\xi} = (\nabla\hat{\xi})^{-1}(\xi - \xi^{*}) \cap \ker\mathsf{C}$, which is not the subspace of singular zero modes, and such a vector subspace cannot be identified for a mechanism with isolated singularities.

\paragraph{Relation to the shape space.}
Using Eq.~\eqref{sm:eq:mpd_perm} requires a parameterization $\psi(\xi)$ of the shape-space $\Sigma$ and the factor $\det\,(\nabla\psi)\trans\nabla\psi$ is the determinant of the induced metric on $\Sigma$.
This is similar to Eq.~\eqref{eq:mpd_regular}, which is the harmonic marginal density.
In contrast, Eq.~\eqref{eq:mpd_singular} does not make an explicit reference to $\Sigma$ and does not involve any parameterization of its branches.
Equation~\eqref{eq:mpd_singular} acquires the factor $\abs{\det\,{\nabla\hat{\xi}}(\nabla\hat{\xi})\trans}^{-1}$ from Eq.~\eqref{sm:eq:coarea}, which a corollary of the coarea formula.
Also, Eq.~\eqref{sm:eq:mpd_perm} is valid for all values of $\xi$, whereas Eq.~\eqref{eq:mpd_singular} is only valid when $\xi$ is close to a singular value $\xi^{*}$ of the CV.

\paragraph{Convergence.}
If a permanently singular mechanism becomes second-order rigid (see Section~\ref{sm:sec:convergence}) once the zero modes are restricted to the subspace $\mathcal{S}$, then $\bm{\sigma}\cdot\bm{w}(\bm{x}) > 0$ for all $\bm{\sigma}$ and the integral in Eq.~\eqref{sm:eq:mpd_perm} converges~\cite{kallus2017}.
In particular, it will not converge for mechanisms that are rigid, but not second-order rigid in $\mathcal{S}$ (e.g., see the example in Appendix A.3 of Ref.~\cite{connelly1996}).
This should be contrasted with the case of Eq.~\eqref{eq:mpd_singular} where there would always be vectors $\bm{t}$ in the integration domain $\Xi_{\xi}$ that satisfy $\bm{\sigma}\cdot\bm{w}(\bm{t}) = 0$ (each corresponding to a tangent to the branch at the singularity).
Hence, convergence of Eq.~\eqref{eq:mpd_singular} relies on the requirement that the number of such vectors is finite. Two necessary conditions required for this are (i) the tangents $\bm{t}$ are resolvable at second order and (ii) the CV map is such that $(\nabla\hat{\xi})\bm{t} \neq \bm{0}$ for all $\bm{t}$ (see Section~\ref{sm:sec:convergence}).

\paragraph{Scaling.}
The scaling of Eq.~\eqref{sm:eq:mpd_perm} with respect to $\beta$ is consistent with Eq.~(9) of Ref.~\cite{kallus2017}.
Furthermore, since the scaling is independent of the value of the CV $\xi$, one would generically expect the free-energy barriers in a permanently singular mechanism to be dominated by entropic effects and the landscape would have the same appearance for all values of $\beta$ (provided it is large).
Contrast this with Eq.~\eqref{eq:mpd_singular}, where the term in the exponent is an inhomogeneous quartic polynomial in the components of $\bm{u}$ for $\xi \neq \xi^{*}$ (see Section~\ref{sm:sec:scaling}).
This makes the scaling nontrivial and gives rise to temperature-dependent barriers in the free-energy landscape.

Returning to the example mechanism in Fig.~\ref{sm:fig:trimer}(a), using Eq.~\eqref{sm:eq:mpd_perm}, we find the marginal density $\mathcal{P}_{\hat{\theta}}(\theta)$ and the free-energy difference $\Delta\mathcal{A}_{\hat{\theta}}(\theta) = -\beta^{-1}\ln\mathcal{P}_{\hat{\theta}}(\theta) + \beta^{-1}\ln\mathcal{P}_{\hat{\theta}}(0)$ to be
\begin{equation}
  \mathcal{P}_{\hat{\theta}}(\theta) = \Gamma\left(\tfrac{1}{4}\right)\left|\frac{2\pi^{6}a^{10}}{3(\beta\kappa)^{7}}\right|^{1/4},
  \quad
  \Delta\mathcal{A}_{\hat{\theta}}(\theta) = \frac{7}{4}\beta^{-1}\ln\left[\frac{\kappa(\theta)}{\kappa(0)}\right].
  \label{sm:eq:trimer_free}
\end{equation}
Above, we have allowed for a hypothetical $\theta$-dependent stiffness $\kappa(\theta)$ so that the scaling factor $7/4$ can be verified.
The numerical results in Fig.~\ref{sm:fig:trimer}(b) show excellent agreement with the analytical predictions.
If the stiffness $\kappa$ is a constant, then $\Delta\mathcal{A}_{\hat{\theta}}(\theta)$ vanishes for all values of $\theta$ and the landscape becomes flat.
This should be contrasted with the examples for mechanisms with isolated singularities, where temperature-dependent free-energy barriers exist even for constant stiffness.

\section{Numerical Simulations}
\label{sm:sec:numerics}

For all the mechanisms we consider in this work, we perform our numerical simulations in the lab frame at an inverse temperature $\beta = 10^{4}$ using a central-force potential $\phi_i[\ell_i(\bm{r})] = [\ell_i^{2}(\bm{r}) - \bar{\ell}_i^2]^2/(8\bar{\ell}_i^2)$, which has an absolute minimum at $\ell_i = \bar{\ell}_i$, for $\ell_i \geq 0$.
With this potential, the stiffness $\kappa_{i} = \phi_i''(\bar{\ell}_i) = 1$ for all bars.
Alternatively, any other potential $\phi_{i}(\ell_{i})$ that depends only on the bar lengths and having a minimum at $\ell_{i} = \bar{\ell}_{i}$ can be used.
We find the marginal probability densities of the CV [Eq.~\eqref{eq:mpd}] using histograms obtained from sampling the Boltzmann--Gibbs distribution using the classical Metropolis Monte Carlo algorithm (with an acceptance rate of about 50\% and $\sim 10^{9}$ samples).
The free-energy profile is then found using Eq.~\eqref{eq:free_energy}.
For the triangulated origami, there is an additional need to reject all Monte Carlo moves that lead to face crossings:
\begin{enumerate}
  \item For faces that share an edge [e.g., faces 1--2--6 and 2--3--6 in Fig.~\ref{sm:fig:origami}(a)], a face crossing can be detected by looking for sign changes in the fold angle of the shared fold when it is close to $\pm \pi$.
  \item For faces that do not share an edge, we use a triangle-triangle intersection test~\cite{tropp2006} to check if they intersect.
    Since there are eight such face pairs, to reduce computational costs, we only check this when the origami is sufficiently folded.
\end{enumerate}

The code we use for Monte Carlo simulations and numerical parameterization of the shape spaces is publicly available~\cite{thermmech}.

\end{document}